\documentclass[12pt]{article}

\usepackage{jheppub}
\usepackage{mathtools}
\usepackage[normalem]{ulem}

\usepackage{etoolbox}
\usepackage{cancel}
\usepackage[percent]{overpic}
\usepackage{xcolor}

\include{defs}

\newtoggle{editing}

	\togglefalse{editing}

\iftoggle{editing}{%
	\usepackage[draft]{showlabels}

	\newcommand{\JSout}[1]{{\textbf{\textcolor{blue}{\sout{#1}}}}}
	\newcommand{\JSeqout}[1]{{\mathbf{\textcolor{red}{\cancel{#1}}}}}
    \newcommand{\JSc}[1]{{\textbf{\textcolor{cyan}{#1  --JS}}}}

}{%

	\newcommand{\JSout}[1]{}
	\newcommand{\JSeqout}[1]{}
    \newcommand{\JSc}[1]{}

}

\begin{document}

\subheader{SU-ITP-15/18 \\ SLAC-PUB-16292}
\title{Tensor Networks from Kinematic Space}
\author[a]{Bart{\l}omiej Czech,} 
\author[a]{Lampros Lamprou,}
\author[a]{Samuel McCandlish,}
\author[b]{James Sully}
\affiliation[a]{Stanford Institute for Theoretical Physics, Department of Physics, Stanford University, Stanford, CA 94305, USA}
\affiliation[b]{SLAC, Stanford University, Menlo Park, CA 94025, USA}
\emailAdd{czech@stanford.edu}
\emailAdd{llamprou@stanford.edu}
\emailAdd{samsamoa@stanford.edu}
\emailAdd{jsully@slac.stanford.edu}

\abstract{
We point out that the MERA network for the ground state of a 1+1-dimensional conformal field theory has the same structural features as kinematic space---the geometry of CFT intervals. In holographic theories kinematic space becomes identified with the space of bulk geodesics studied in integral geometry. We argue that in these settings MERA is best viewed as a discretization of the space of bulk geodesics rather than of the bulk geometry itself. As a test of this \emph{kinematic proposal}, we compare the MERA representation of the thermofield-double state with the space of geodesics in the two-sided BTZ geometry, obtaining a detailed agreement which includes the entwinement sector. We discuss how the kinematic proposal can be extended to excited states by generalizing MERA to a broader class of compression networks.}
\maketitle

\section{Introduction}

Gravitational physics presents us with a paradox. On the one hand, its most successful formulation to date -- the general theory of relativity -- relies on differential geometry, which emphasizes local dynamics. On the other hand, all gauge-invariant observables in gravity live on the asymptotic boundary and are therefore global in character. While the local approach has been pursued with undiminished success for one century \cite{einstein1915}, a more global strategy has not yet congealed into a unified formalism. The best developed attempt to fill this gap is the AdS/CFT correspondence \cite{adscft}, which organizes the gauge-invariant quantities in a gravitational spacetime into a field theory living on its asymptotic boundary. A key challenge facing the holographic program -- one that has come to the spotlight in recent years \cite{samirstheorem, amps} -- is this: how do we reconcile the CFT-based, global formulation of gravity with the local language of general relativity?

In Ref.~\cite{lastpaper}, we outlined a semantically evident answer to this question: to complement Einstein's apparatus of differential geometry, we need an approach based on integral geometry \cite{santalo}. This beautiful field of mathematics is concerned with translating between local and global properties of geometric spaces. A well-known application is to recover a function from its integrals along straight lines \cite{helgason}, a problem that occurs in seismology \cite{seismology}, medical imaging \cite{ctscans} and the reconstruction of bulk operators in holographic duality \cite{ooguri}. In \cite{lastpaper}, we focused on a prequel to this problem: determining the geometry of an asymptotically AdS$_3$ spacetime from data in the dual conformal field theory.

Generally, the input to the reconstruction problem consists of all correlation functions in the CFT. However, recent years have taught us that information-theoretic CFT data are particularly robust probes of the bulk geometry. The foremost among them are entanglement entropies of boundary regions, which compute areas of bulk minimal surfaces \cite{rt1, rt2, hrt}.\footnote{Numerous other gravitational quantities also have information-theoretic dual descriptions, including a version of the null energy condition \cite{nec}, Einstein's equations \cite{einsteineq1, einsteineq2, einsteineq3}, canonical energy \cite{canenergy}, lengths of curves \cite{holeography, robproof, protocol}, the triangle inequality \cite{lampros} and even connectedness of spacetime \cite{marksessay, mark2}.} In Ref.~\cite{lastpaper}, we used entanglement entropies to define an auxiliary, Lorentzian geometry, whose points are in one-to-one correspondence with boundary intervals and, by the Ryu-Takayanagi proposal \cite{rt1}, with spacelike geodesics in the dual geometry. The resulting object, called kinematic space, is an intermediary in the AdS/CFT translation, providing a natural volume form on the space of bulk geodesics. Integrals of that form compute lengths of all bulk curves in a generalization of the famous Crofton formula, which tells us how likely a dropped needle is to land on a single bathroom tile \cite{bouffon}. 

But the problems solved by kinematic space are not confined to holographic duality. Even in the absence of a gravitational dual, $d$-dimensional conformal field theory intertwines space and scale (RG direction), as is evident from its global symmetry group SO$(d,2)$. One may imagine that kinematic space, which organizes the entanglement structure of a state by location and scale, may have already found use in the study of conformal field theories, independently of holographic considerations. If so, in what form has kinematic space previously appeared?

The answer is the Multi-Scale Entanglement Renormalization Ansatz (MERA) \cite{mera, mera2}. The present paper explains the merits of viewing the MERA network as a discrete version of the vacuum kinematic space. 
The argument makes crucial use of the auxiliary causal structure of MERA, which originates from working with unitary and isometric tensors as part of the ansatz. 
This causal structure was independently exploited to argue that MERA most naturally lives on de Sitter space \cite{beny}.
Our key insight is to recognize that this de Sitter space is the vacuum kinematic space, which carries an information metric determined by entanglement.  
This allows us to propose a generalization of MERA to excited states \cite{compression}.

\begin{figure}
\centering 
\includegraphics[width=0.6\textwidth]{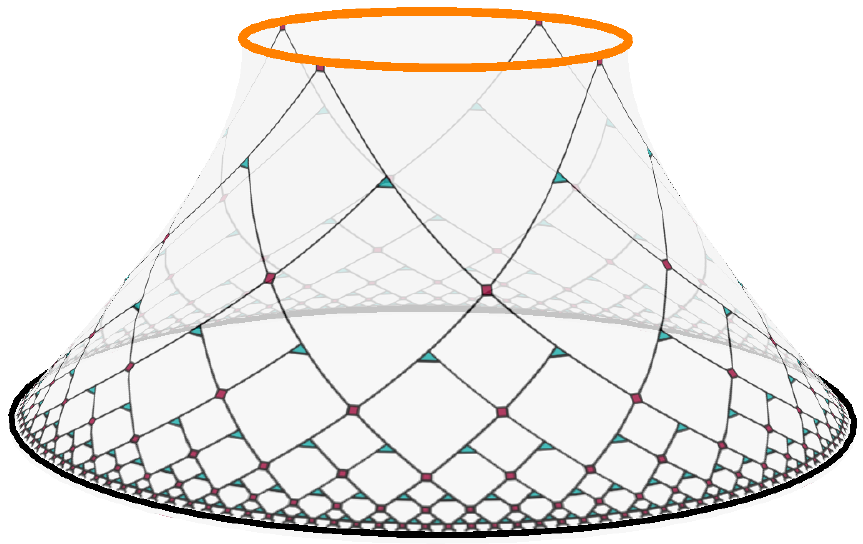} 
\caption{MERA naturally lives on half of two-dimensional de Sitter space, the kinematic space for an equal-time slice of AdS$_3$.}
\label{fig:MERAondS} 
\end{figure}

Our results do not preclude other types of networks, for example ones studied in \cite{errorcorrectingnetwork, patrickxiaoliang}, from discretizing a time slice of the dual geometry directly. 

\paragraph{Reading guide} In an effort to keep the paper self-contained, we begin with a review of integral geometry and MERA (Sec.~\ref{sec:review}). 
Secs.~\ref{sec:therelation} and \ref{sec:BHmera} contain the main arguments for identifying MERA with the kinematic geometry. Sec.~\ref{sec:therelation} is set in the broad context of the ground state MERA while Sec.~\ref{sec:BHmera} discusses the recently reported \cite{quotientmera} MERA construction of the thermofield double state. 
Sec.~\ref{compressionsummary} briefly motivates the results of the second part of this work, which will be presented in \cite{compression}. 
Sec.~\ref{summary} summarizes our core arguments and discusses the main results of this work.

\section{Review}
\label{sec:review}
We begin by reviewing the properties of kinematic space and the MERA tensor network. The reader is encouraged to look for commonalities.

\subsection{Kinematic space}
A more complete discussion of the ensuing material was given in \cite{lastpaper}.

\subsubsection{Crofton's formula in flat space}
Crofton's formula states that the length of a curve is measured by the number of straight lines that intersect it. To state this result formally, we need to clarify how to count straight lines. Straight lines on the plane form a two-dimensional manifold $K$ known as `kinematic space.' To quantify `how many' straight lines $g$ satisfy some condition, we need a homogeneous measure $\mathcal{D}g$ on kinematic space. Using translations and rotations fixes the measure, up to a multiplicative constant, to be
\begin{equation}
\mathcal{D}g = dp \wedge d\theta.
\label{flatmeasure}
\end{equation}
Here $p$ is the distance of the straight line from the origin and $\theta$ is the angle it makes with some fixed axis. Allowing $p$ to be negative extends the measure to the set of oriented straight lines. 

Crofton's formula states that, for every curve $\gamma$ of finite length,
\begin{equation}
\textrm{length of }\gamma = \frac{1}{4}
\int_K n(g,\gamma)\,\,\mathcal{D}g.
\label{croftonformula}
\end{equation}
Here $n(g, \gamma)$ is the number of intersections between the straight line $g$ and $\gamma$. This result can be used to solve Buffon's needle problem \cite{bouffon}, which we referenced in the Introduction. Our primary interest is in an extension of this formula to holographic spacetimes.

\subsubsection{Crofton's formula in holographic geometries}

\begin{figure}
\centering 
\includegraphics[width=0.3\textwidth]{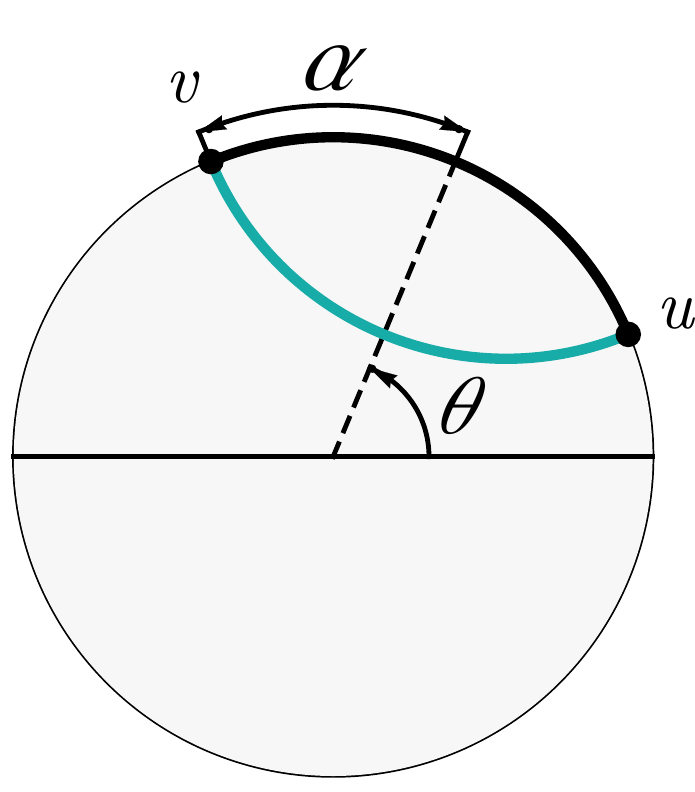} 
\:\:\:\:\:\:\:\:\:\:\:\:\:\:\:\:\:\:\:\:\:\:\:\:  
\includegraphics[width=0.30\textwidth]{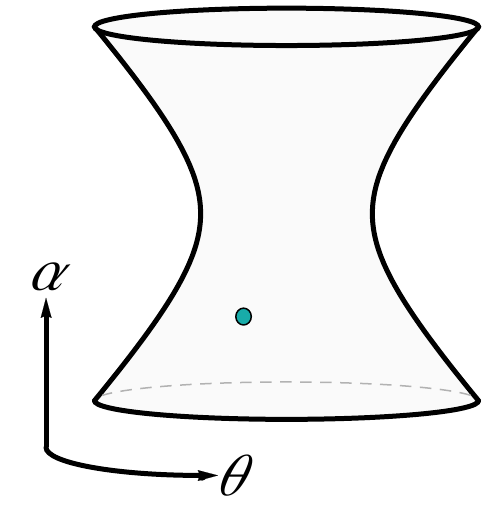} 
\caption{The kinematic coordinates $\alpha$ and $\theta$ correspond to the
half-opening angle of the geodesic and the angular location of its
center-point repectively. Geodesics in the hyperbolic plane are mapped to points on kinematic space.}
\label{fig:KScoords} 
\end{figure}

To extend eq.~(\ref{croftonformula}) to a static slice of a holographic
geometry, one need only supply the correct measure on the generalized kinematic space. In pure AdS$_{3}$
of curvature radius $L$, the measure is again uniquely selected by
invariance under the isometries of $\mathbb{H}_2$ \cite{solanes}, and takes the form 
\begin{equation}
\mathcal{D}g=\frac{L\,d\theta\wedge d\alpha}{\sin^{2}\alpha}.\label{hypmeasure}
\end{equation}
This expression gives a `density of geodesics' near a geodesic centered
at a boundary point $\theta$, with opening angle $\alpha$ (see Fig.~\ref{fig:KScoords}). Looking ahead to a connection with the MERA
network, we note that $\theta$ and $\alpha$ play the role of position
and scale in the CFT, respectively.

In a general, static, holographic geometry, the measure cannot be
found from symmetry alone. In Ref. \cite{lastpaper}, we
showed that when the tangent bundle of a bulk time-slice is covered by boundary-anchored
geodesics, the measure takes a remarkably simple form,
\begin{equation}
\mathcal{D}g=\frac{\partial^{2}S(u,v)}{\partial u\,\partial v}\,du\wedge dv,\label{kinvolume}
\end{equation}
where $S(u,v)$ is the length of a geodesic connecting the boundary
points $u$ and $v$. Here, we have defined `light-cone' coordinates
on kinematic space,
\begin{equation}
u=\theta-\alpha\qquad{\rm and}\qquad v=\theta+\alpha,\label{nullcoords}
\end{equation}
which label a geodesic by its asymptotic endpoints. Then eq. (\ref{kinvolume})
agrees with eq.~(\ref{hypmeasure}) after substituting $S(u,v)=2L\log\sin\frac{v-u}{2}+{\rm const}$.

When $S(u,v)$ refers to the length of the shortest geodesic connecting points $u$ and $v$ satisfying the homology condition, its length in units of $4G$ is the entanglement entropy of the boundary interval $(u,v)$ \cite{rt1}. Thus, it is convenient to divide both sides of eq.~(\ref{croftonformula}) by $4G$ and obtain:
\begin{equation}
\frac{\textrm{length of }\gamma}{4G} = \frac{1}{4}
\int_K n(g,\gamma)\,
\frac{\partial^2 S_{\rm ent}(u,v)}{\partial u \, \partial v}\, du\wedge dv
\label{croftonformula4G}
\end{equation}
In what follows, we set $4G \equiv 1$ and do not distinguish between $S(u,v)$ as a length and $S_{\rm ent}(u,v)$ as an entanglement entropy.

\paragraph{Differential entropy} For a closed curve, it is instructive to carry out the integral in (\ref{croftonformula4G}) explicitly in one direction:
\begin{equation}
\textrm{length of }\gamma = 
- \int_0^{2\pi} du\, 
\frac{\partial S(u,v)}{\partial u}\,\Big|_{v = v(u)} = S_{\rm diff}
\label{sdiff}
\end{equation}
This expression is the differential entropy, first reported in \cite{holeography} (see also \cite{robproof}). It localizes on the set of geodesics tangent to $\gamma$, which is the boundary of the set of its intersecting geodesics. The tangency condition appears through $v(u)$, which is defined by demanding that the geodesic connecting $u$ and $v(u)$ be tangent to $\gamma$.

In Sec.~\ref{cutcount} we will find an analogue of this expression in the cut-counting prescription for estimating entropies in MERA.

\subsubsection{Causal structure and the kinematic metric}

\label{kincausality} The kinematic space for a static slice of an
asymptotically AdS$_{3}$ geometry has a richer structure than just
a density form: It can also be equipped with a metric with mixed signature.
To see this, note that the space of geodesics maps naturally to the
space of boundary intervals via the R-T prescription. The causal structure
of kinematic space descends not from the causal structure of AdS$_{3}$,
but from the partial ordering of boundary intervals by containment.

In particular, given two boundary intervals $A,B$ corresponding to
two points $a,b$ in kinematic space, we say that $a$ causally precedes
$b$ if $A\subset B$. Any pair of geodesics may then be classified
as timelike, lightlike, or spacelike-separated:
\begin{itemize}
\item Timelike: geodesic $(u_{1},v_{1})$ is said to live in the past of
geodesic $(u_{2},v_{2})$ if 
\begin{equation}
[u_{1},v_{1}] \subset [u_{2},v_{2}]\label{containment}
\end{equation}
as intervals on the asymptotic boundary. Note that the direction of
kinematic `time' reverses under changes of orientation. For the same
geodesics with opposite orientation, we have $[v_{2},u_{2}]\subset[v_{1},u_{1}]$. 
\item Spacelike: geodesics $(u_{1},v_{1})$ and $(u_{2},v_{2})$ are spacelike
separated when neither interval contains the other. 
\item Lightlike: This is the borderline case between spacelike and timelike
separation. It occurs when one of the intervals subtended by the geodesics
contains the other, but only marginally. This means that the intervals
share an endpoint -- on the left or on the right: 
\begin{equation}
u_1=u_2 \,\,\,\text{
or }\,\,\, v_1=v_2
\end{equation}
 This is the reason why above eq.~(\ref{nullcoords})
we referred to the endpoint coordinates of kinematic space as `light-cone'
coordinates. 
\end{itemize}

\paragraph{Kinematic Metric}

For a time-slice of pure AdS$_{3}$, we can now see that symmetry
fixes the metric on kinematic space to be the
two-dimensional de Sitter metric \cite{solanes,lampros}:
\begin{equation}
ds_{{\rm kin}}^{2}=\frac{L}{\sin^{2}\alpha}\left(-d\alpha^{2}+d\theta^{2}\right).\label{adskinmetric}
\end{equation}
To see this, note that $\text{dS}_{2}$ is the only metric space with
$\text{SO}\left(2,1\right)$ isometry group that realizes the requisite causal
structure. With the coefficient above, the volume
form $d^{2}V_{\text{kin}}$ in kinematic space is equal to the geodesic
density $\mathcal{D}g$ of eq.~(\ref{hypmeasure}).

Moving to a general holographic geometry, specifying the causal structure
and the volume form $d^{2}V_{\text{kin}}=\mathcal{D}g$ yields a unique
Lorentzian metric: 
\begin{equation}
ds_{{\rm kin}}^{2}=\frac{\partial^{2}S(u,v)}{\partial u\,\partial v}\,du\,dv\label{kinmetric}
\end{equation}
The relevance of this metric for reconstructing local features
of the bulk geometry was reported in \cite{lampros,lastpaper}.

\subsubsection{Conditional mutual information in kinematic space}
\label{kincmi}

\begin{figure}
	\centering
	\includegraphics[width=0.60\textwidth]{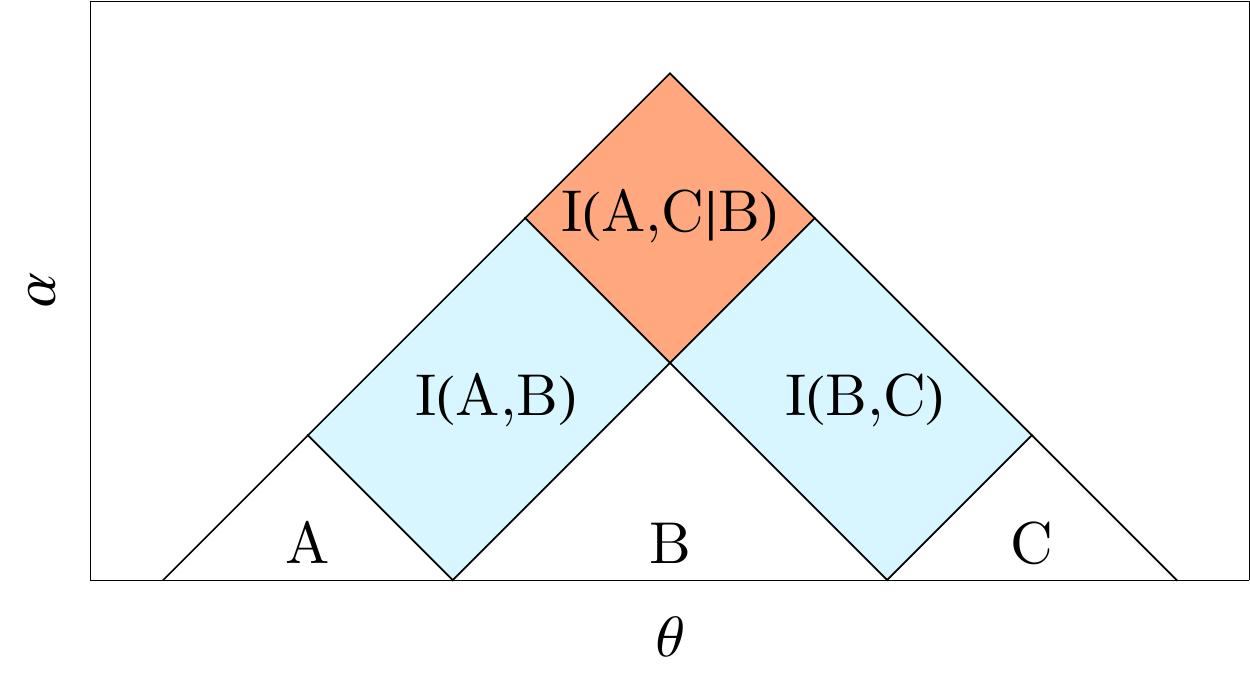}
	\caption{Volumes of causal diamonds in kinematic space compute conditional mutual informations of triples of contiguous intervals. As a special case, causal diamonds with one vertex on the boundary compute mutual informations of adjacent intervals.}
	\label{fig:KSmutual}
\end{figure}

The volume form~(\ref{kinvolume}) has a meaning in information theory. Conditional mutual information is defined as the following combination of entanglement entropies:
\begin{equation}
I(A,C|B) = S(AB) + S(BC) - S(ABC) - S(B)
\label{defcmi}
\end{equation}
Mutual information $I(A, C)$ is a special case of this quantity, conditioned on \mbox{$B = \emptyset$}. Conditional mutual information is also familiar from the strong subadditivity of entanglement entropy, which guarantees that it is non-negative \cite{ssa}. For the special choice
\begin{equation}
A = (u-du, u) \qquad {\rm and} \qquad
B = (u, v) \qquad {\rm and} \qquad
C = (v, v+dv)\,,
\label{3intervals}
\end{equation}
we have:
\begin{equation}
S(u-du,v) + S(u,v+dv) - S(u-du,v+dv) - S(u,v) = 
\frac{\partial^2 S(u,v)}{\partial u \,\partial v}\,du\,dv = d^2V_{\rm kin}\,. 
\label{infinitesimalcmi}
\end{equation}
Eq.~(\ref{infinitesimalcmi}) states that the Lorentzian area of an infinitesimal causal diamond in kinematic space computes the conditional mutual information of a triple of neighboring intervals  (\ref{3intervals}). Owing to the chain rule for conditional mutual information
\begin{equation}
I(A, CD|B) = I(A,C|B) + I(A, D|BC)\,,
\end{equation}
this conclusion automatically extends to all causal diamonds in kinematic space, regardless of size (see Fig.~\ref{fig:KSmutual}). A special case is a causal diamond with one of its vertices on the boundary of kinematic space, whose volume is equal to the mutual information of the two adjacent intervals. Thus, eq.~(\ref{croftonformula4G}) states that the length of any curve on a static slice of an asymptotically AdS$_3$ geometry computes a combination of conditional mutual informations.

\subsection{The MERA network}
Our presentation will be brief, because good reviews exist elsewhere \cite{mera-review, mera-revresults}. We highlight those aspects of MERA, which are key for appreciating the connection with kinematic space. A reader familiar with MERA may skip over to Sec.~\ref{sec:therelation}.

\subsubsection{Tensor network generalities}
The wavefunction of a general $N$-body system defines a tensor with $N$ indices:
\begin{equation}
| \Psi \rangle = \sum_{i_1i_2\ldots i_N} 
\Psi_{i_1i_2\ldots i_N}|i_1 i_2 \ldots i_N\rangle
\label{genwavefn}
\end{equation}
The walloping number of components of this tensor --exponential in $N$-- reflects the complexity of an arbitrary many-body wavefunction. However, imposing physical constraints such as locality and symmetry ought to simplify the description of the wavefunction drastically. This simplification is the objective of tensor network techniques. 

\begin{figure}
	\centering
	\includegraphics[width=0.95\textwidth]{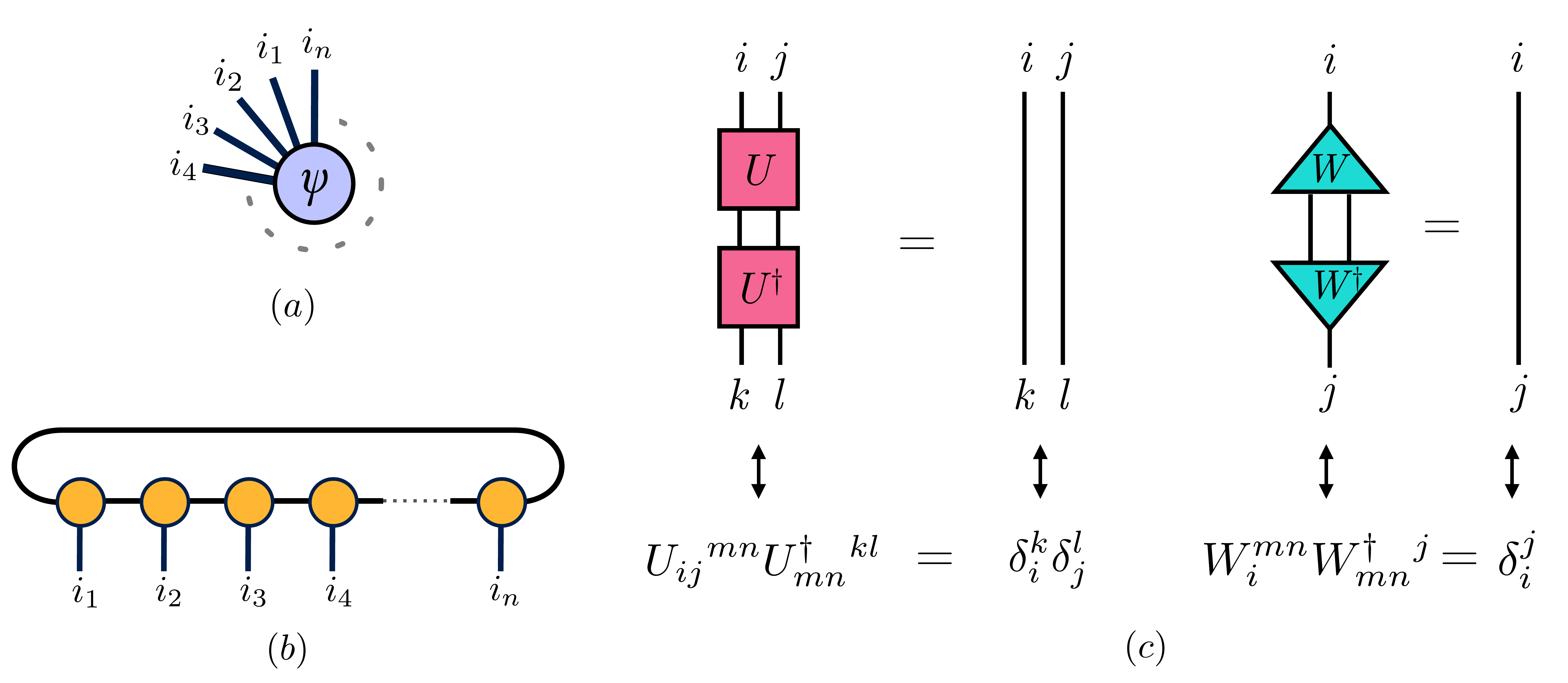}
	\caption{Examples of tensor networks. (a) A featureless tensor network composed of a single tensor. This can prepare a generic state, as in eq.~(\ref{genwavefn}). (b) A tensor network composed of a chain of tensors contracted together (a matrix product state). (c) The unitary (resp. isometric) character of the disentanglers and isometries in MERA means that these tensors cancel out when contracted with their hermitian conjugates.}
	\label{fig:tensors}
\end{figure}

Tensor networks are graphs, which consist of vertices and edges. Every vertex stands for a tensor with as many indices, as there are edges incident on it. The indices range from 1 to $\chi$, the `bond dimension' of a given edge. An edge connecting two vertices denotes a common index of two tensors, which is contracted (traced out.) Some examples of tensor networks, including the featureless wavefunction from eq.~(\ref{genwavefn}), are shown in Fig. \ref{fig:tensors}.   

\subsubsection{Structure of the MERA network}
\label{merastructure}
\begin{figure}
	\centering
	\includegraphics[width=0.60\textwidth]{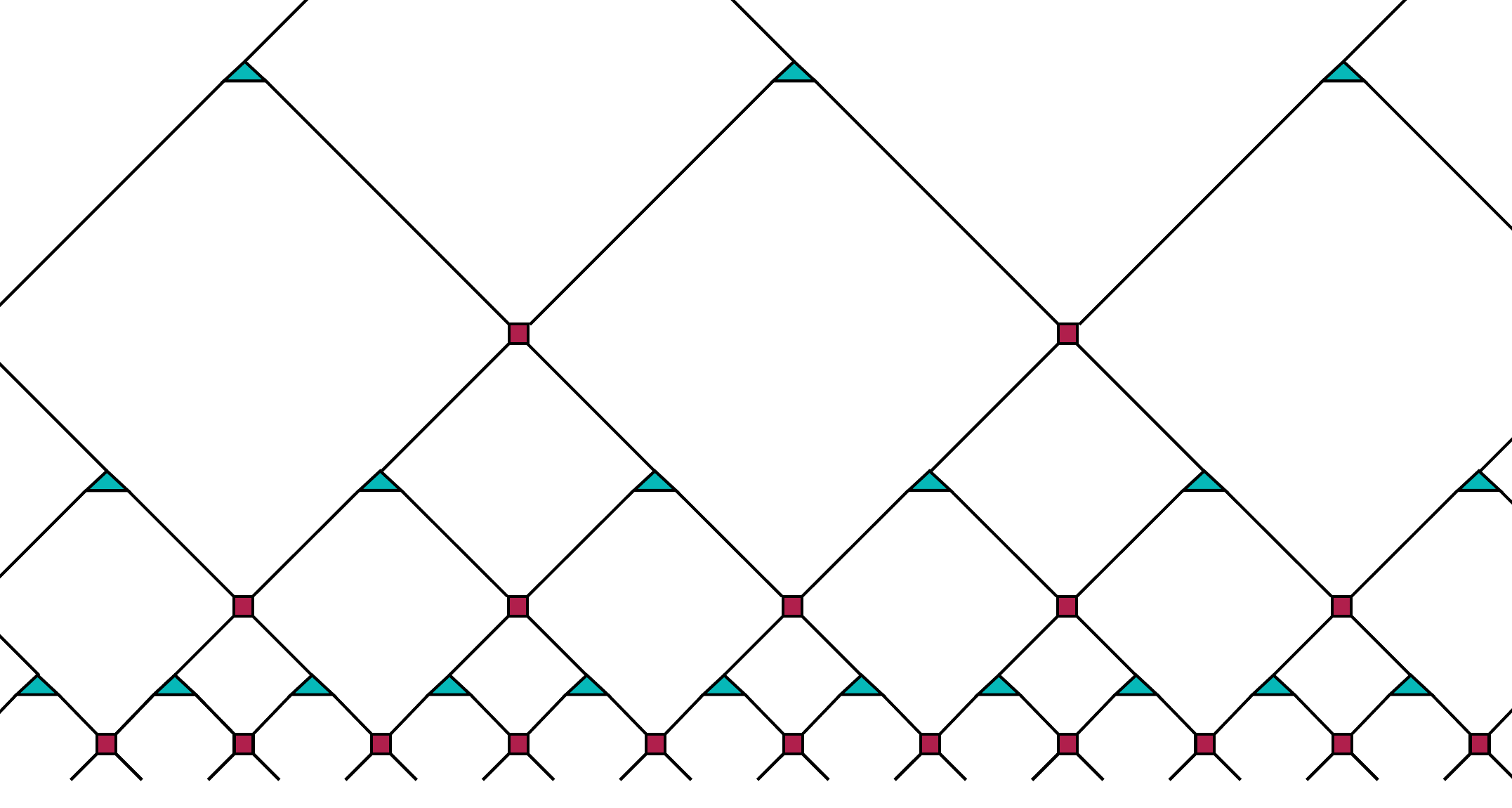}
	\hfill
	\includegraphics[width=0.34\textwidth]{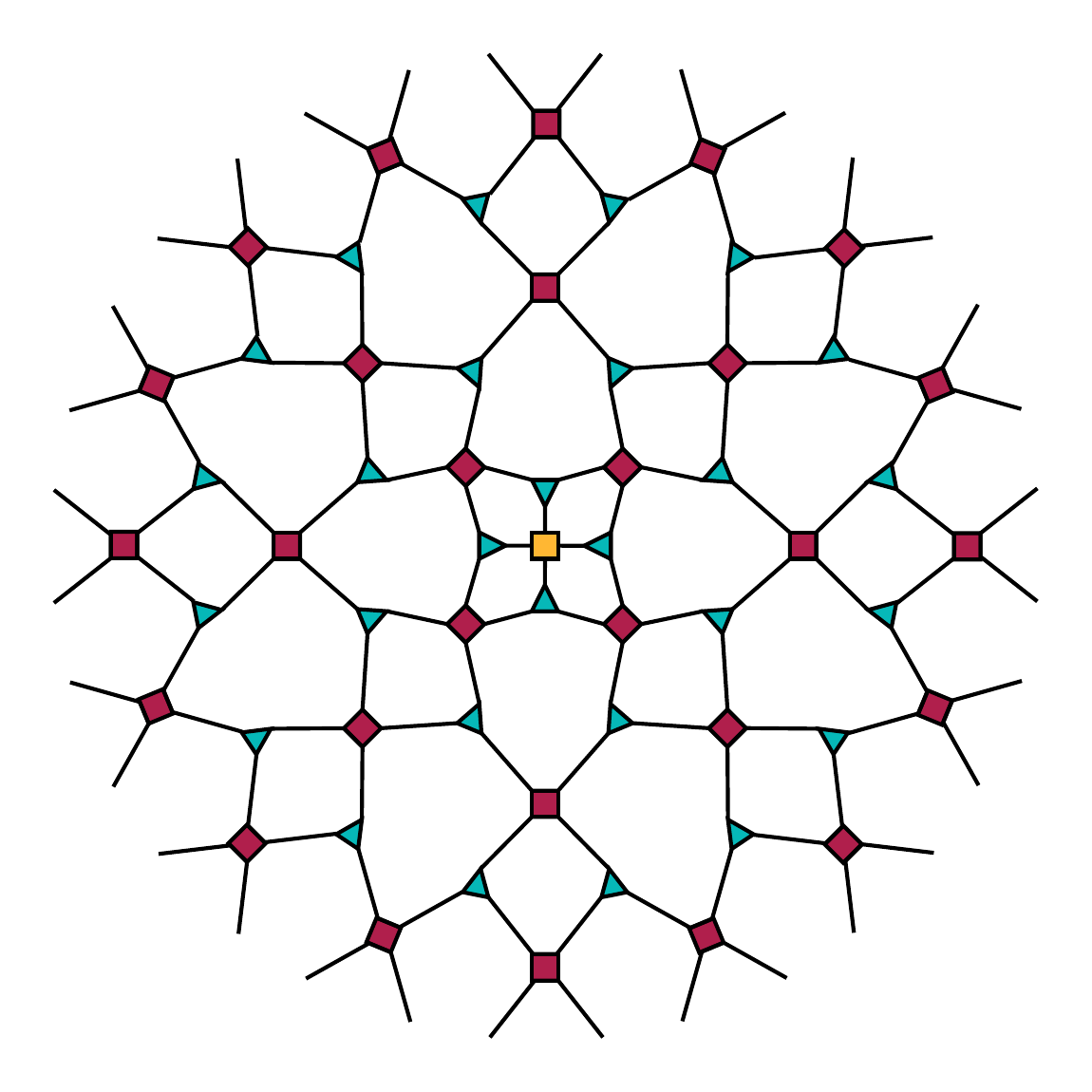}
	\caption{The MERA lattices for states on a line and a circle.}
	\label{fig:mera}
\end{figure}

The MERA network is a successful ansatz for the ground state wavefunction of a conformal field theory \cite{mera-revresults}. For a CFT$_2$---the case of interest in the present paper---it is a two-dimensional array of tensors shown in Fig.~\ref{fig:mera}. While the horizontal direction corresponds to the spatial axis of the CFT, the vertical direction is meant to encode scale (RG direction). In a true CFT, which has no characteristic scale, the vertical direction ought to be infinite. In practice, however, MERA networks are presented with a finite number of layers, which is tantamount to fixing a UV cutoff. 

Because the axes of MERA correspond to space and scale, the network provides a graphical representation of renormalization in real space. For example, cutting the network one layer higher takes the wavefunction at scale $\mu$ to the wavefuction at the coarser scale $2\mu$. More generally, cutting the network in an inhomogeneous way can be understood as enacting a local scale transformation \cite{quotientmera}. In this way, the hierarchical structure of MERA encodes an iterative application of local coarse-graining transformations. To understand the rationale underlying the MERA ansatz, it is useful to examine a single layer of the network and ask how it is intended to coarse-grain the wavefunction.

\paragraph{Disentanglers and isometries} A layer of MERA consists of two types of tensors laid out in two rows. The tensors with four legs are called disentanglers. They are $\chi^2 \times \chi^2$ unitary transformations, which select bases wherein incoming UV degrees of freedom will appear locally unentangled. This change of basis is performed in order to prevent UV entanglement from accumulating in the IR wavefunctions defined on higher cuts. In this way, through the action of disentanglers, the MERA network partitions entanglement entropies of intervals into scale-specific contributions. 

The second component in every layer of MERA is a row of isometries. These project the locally disentangled UV degrees of freedom into the effective IR Hilbert space. This transformation is isometric, which means that it can be extended to a unitary map 
\begin{equation}
\mathcal{H}_{\rm UV} \to \mathcal{H}_{\rm IR} \otimes \mathcal{H}_{\rm frozen}\,,
\label{frozen}
\end{equation}
with $\mathcal{H}_{\rm frozen}$ not participating in further steps of the renormalization \cite{mera2, xiaoliang}. Our diagrams ignore $\mathcal{H}_{\rm frozen}$, showing isometries as maps from two UV lines (a $\chi^2$-dimensional vector space) to a single IR line (a $\chi$-dimensional vector space.)

Of course, not every wavefunction can be prepared with this ansatz. This is the price we pay for efficiency---by varying the tensors in this fixed network, we scan an $\mathcal{O}(\chi^4 N)$-dimensional corner of the full Hilbert space, which we hope includes the ground state wavefunction. This hope has been validated in numerous computations, with the optimized MERA (the state of lowest energy in the variational class \cite{mera-opt}) correctly reproducing the spectrum and OPE coefficients of CFT$_2$s such as the critical Ising model \cite{mera-revresults, mera-expl, mera-moredata}.

\subsubsection{Causal structure}
\label{MERAcausality}
This fundamental feature of the MERA network, noticed and exploited already in the initial papers on the subject \cite{mera2, mera-expl}, offers the first hint of a relation to kinematic space. A prescient proposal relating MERA to de Sitter space appeared in \cite{beny}.

Consider the reduced density matrix of an interval $\mathcal{I}$ in a pure state $|\Psi\rangle$:
\begin{equation}
\rho_\mathcal{I}={\rm Tr}_{\mathcal{I}^c}|\Psi\rangle\langle\Psi | \, .
\label{rhoi}
\end{equation}
In the language of MERA, we compute it by putting together the tensor network representations of the bra and ket states and joining (tracing out) indices not contained in $\mathcal{I}$.
Tracing out these indices means that disentanglers from the $|\Psi\rangle$ network get contracted with their hermitian conjugates from the $\langle\Psi |$ network and cancel out (compare with Fig.~\ref{fig:tensors}(c)). A similar cancellation occurs in the isometries above them, then in the next row of disentanglers, and so on. The ensuing cascade of cancellations divides the network into two parts: the region that determines $\rho_\mathcal{I}$, and the region that drops out from it. In analogy with the propagation of signals in a Lorentzian spacetime, we call the former region the `inclusive causal cone' of interval $\mathcal{I}$.

\begin{figure}
	\centering
		\includegraphics[width=0.75\textwidth]{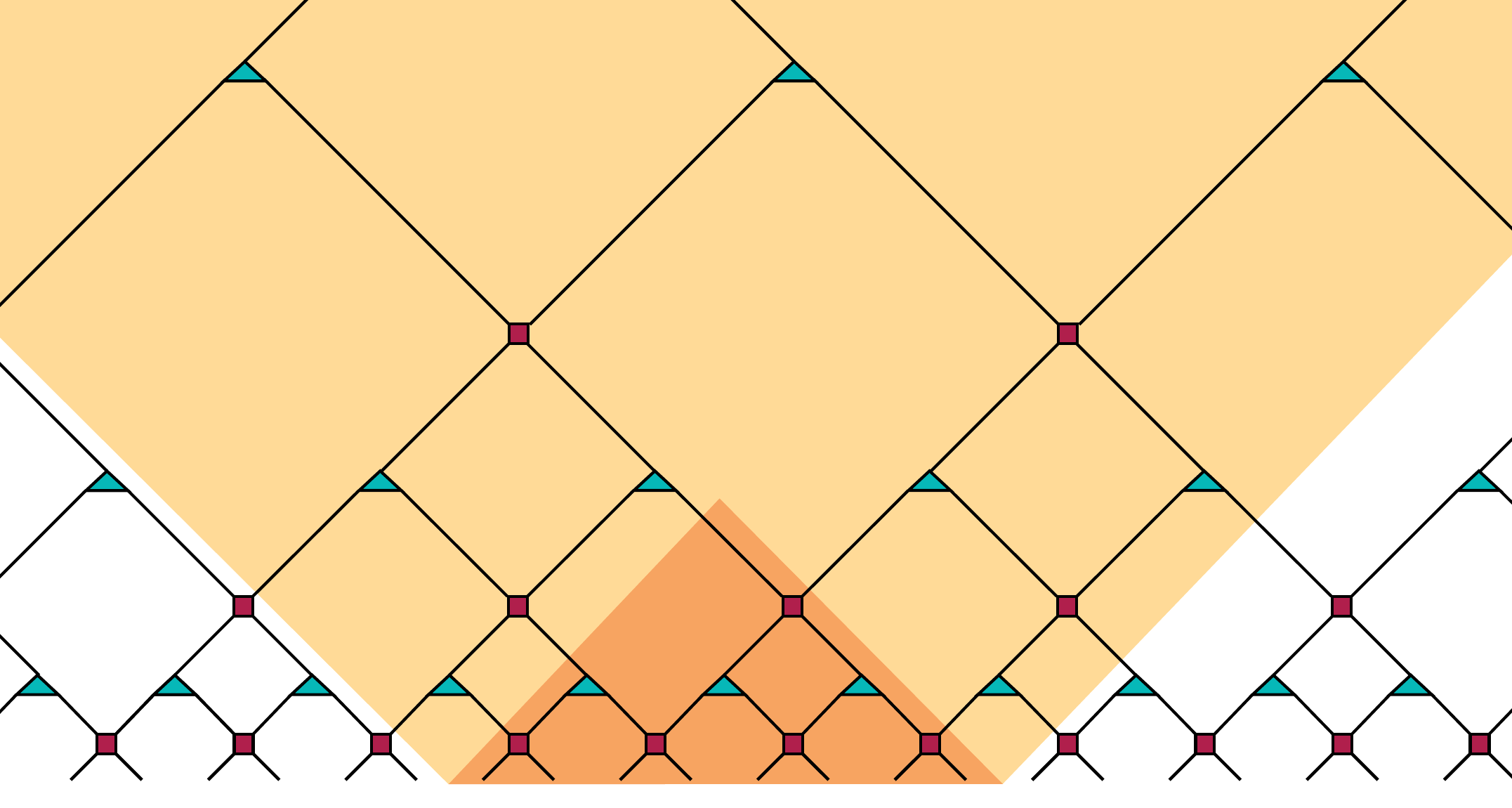}
	\caption{The MERA causal structure. The inclusive causal cone of eight terminal sites is indicated by the expanding yellow region of the network. The exclusive causal cone of the same eight UV sites is indicated by the contracting orange region of the network. The tensors in this region perform a change of basis, which takes the state living on the five sites on the lightlike cut to the eight sites in the UV.}
	\label{fig:cone}
\end{figure}

We shall see in a moment that this notion of `causality' is the same as in Sec.~\ref{kincausality}. Before explaining this, let us consider $\rho_{\mathcal{I}^c}$, the reduced density matrix of the complement of $\mathcal{I}$. It too splits up the MERA network into two regions -- the inclusive causal cone of $\mathcal{I}^c$ and the rest. Altogether, the division of the Hilbert space into localized tensor factors
\begin{equation}
\mathcal{H} = \mathcal{H}_\mathcal{I} \otimes \mathcal{H}_{\mathcal{I}^c}
\label{factors}
\end{equation}
partitions the MERA network into three components: a region that only affects $\mathcal{I}$, an analogous region for $\mathcal{I}^c$, and a region that affects the reduced states of both. This division is shown in Fig.~\ref{fig:cone}. For obvious reasons, the first two regions are often called the exclusive causal cones of their respective intervals.

\begin{figure}
	\centering
		\includegraphics[width=0.75\textwidth]{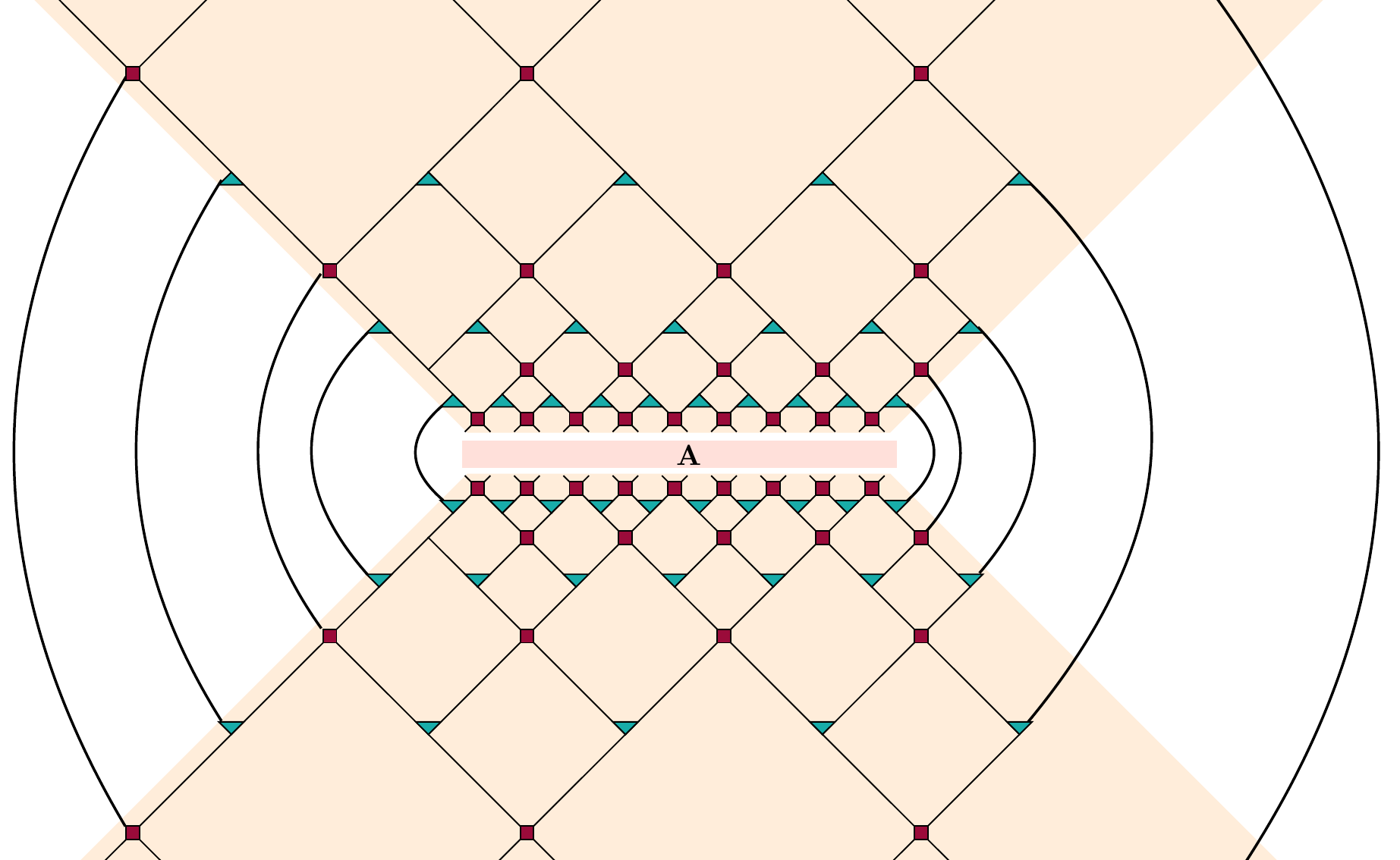}
	\caption{The reduced density matrix $\rho_A$ for an interval $A$ in the vacuum can be represented as a tensor network.  This network is obtained by taking two copies of MERA, then tracing over $A^c$.  This causes a cascade of cancellations of disentanglers and isometries.  The remaining tensors are only those in the inclusive causal cone of $A$.}
	\label{fig:density-matrix}
\end{figure}

\paragraph{Lightlike coordinates} To keep the nomenclature consistent, we ought to call the boundaries of the aforementioned regions `lightlike.' These lightlike directions -- one left-going and one right-going -- are linear combinations of the two axes of MERA:
\begin{equation}
{\rm lightlike} = {\rm location}\pm {\rm scale}
\quad \leftrightarrow \quad u, v~\textrm{from eq. }(\ref{nullcoords}) \, .
\label{MERAcoords}
\end{equation}
As in kinematic space, it is convenient to use them as coordinates on MERA. Doing so canonically assigns an interval to every tensor in the network. Specifically, a tensor at lightlike coordinates $u$ and $v$ is the topmost component of the exclusive causal cone of interval $(u,v)$. Notice that the notion of causality defined by eq.~(\ref{MERAcoords}) in MERA is exactly the same as that in Sec.~\ref{kincausality}: if a tensor at $(u_2, v_2)$ is in the MERA-future (past) of the tensor at $(u_1, v_1)$, the corresponding interval contains (is contained in) its counterpart.

The privileged role of the lightlike directions in MERA is a consequence of working with unitary tensors; without unitarity, cancellations discussed below eq.~(\ref{rhoi}) would not occur and all parts of the network would affect $\rho_{\mathcal{I}}$ and $\rho_{\mathcal{I}^c}$. This marriage of unitarity and causality is displayed by the exclusive causal cones of intervals, whose role amounts to a change of basis. Observe that the action of tensors in the exclusive causal cone of $\mathcal{I}$ is undetectable by observables in $\mathcal{I}^c$, so it is a transformation within $\mathcal{H}_\mathcal{I}$. After concatenating with the exclusive causal cone, the rather abstract state defined on its lightlike edges is mapped into a local basis of $\mathcal{H}_\mathcal{I}$. Although the linear map effected by the exclusive causal cone is an isometric embedding of a smaller Hilbert space in a larger one, when the frozen degrees of freedom from eq.~(\ref{frozen}) are taken into account, it is manifestly unitary. 


\subsubsection{Entanglement entropies from cut-counting} 
\label{cutcount}
\begin{figure}
	\centering
	\includegraphics[width=0.7\textwidth]{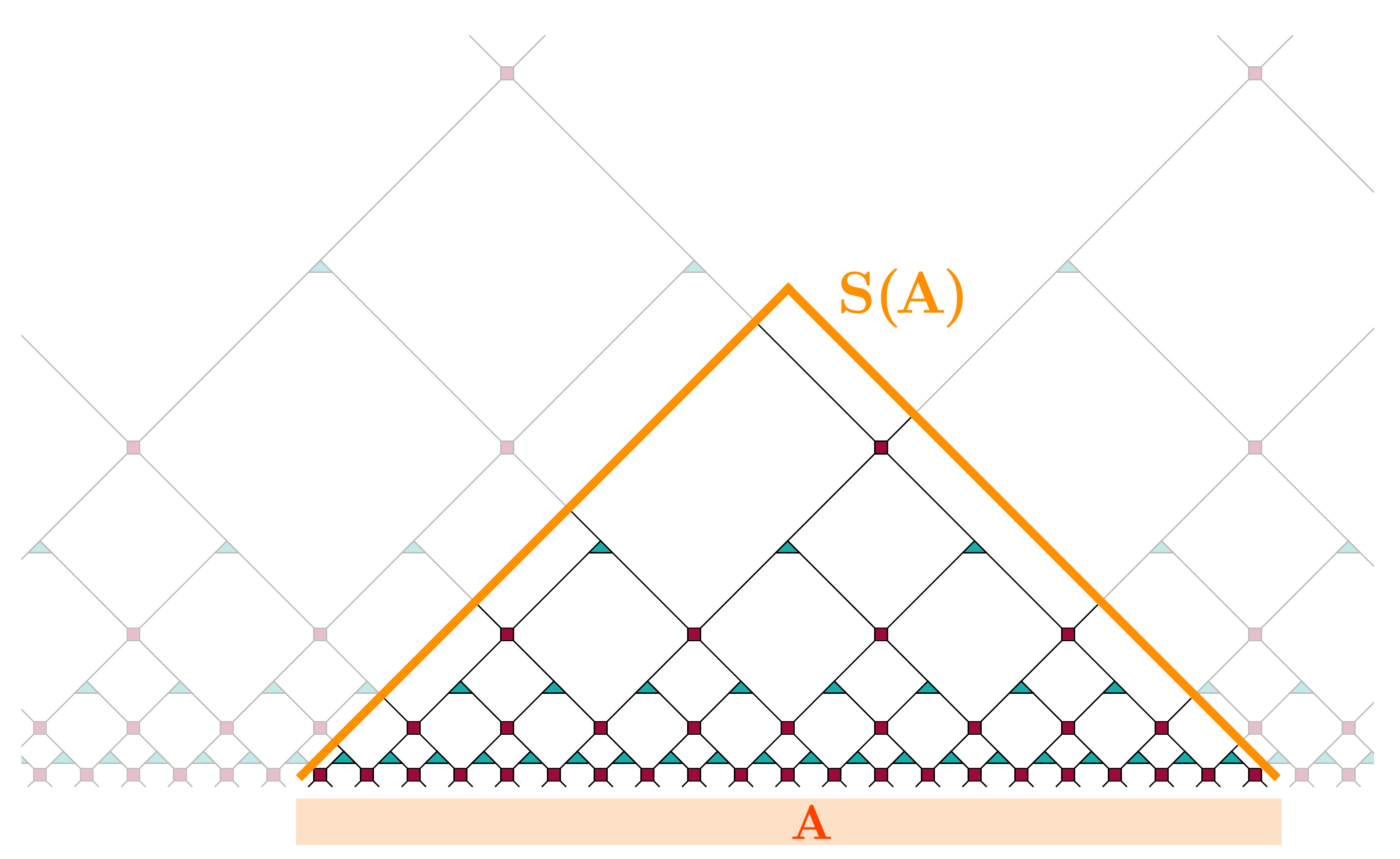}
	\caption{Up to an overall coefficient, the entanglement entropy of a CFT interval $A$ can be estimated by the number of network lines crossing the exclusive causal cone of $A$ times the logarithm of the bond dimension $\log \chi$.}
	\label{fig:cutcounting}
\end{figure}

A central motif of the present work---and one that motivated holographers' initial interest in MERA \cite{briansessay}---is the simple way the network encodes entanglement entropies. For intervals of less than half system size, a good estimate is obtained by counting the number of lines emanating from the exclusive causal cone of the interval. If each line is counted with weight $\log \chi$, this amounts to computing the logarithm of the dimension of the Hilbert space living on the edge of the exclusive causal cone; see Fig.~\ref{fig:cutcounting}. In what follows, we will refer to this edge as the `causal cut,' though the term `minimal curve' has been used in prior literature \cite{briansessay, brianspaper}.

On the one hand, the cut-counting prescription gives a manifest upper bound on the entanglement entropy. We saw in Sec.~\ref{MERAcausality} that the spectrum of the reduced density matrix of the interval is prepared above the causal cut. The tensors below the cut merely choose a basis in which the state is expressed and therefore have no effect on the entanglement entropy. The maximal value of the entanglement entropy is the logarithm of the dimension of the Hilbert space, in which the state prepared by the network lives. This is precisely what the cut-counting prescription computes.

Though there is no similar argument bounding the entanglement entropy from below, in practice counting cuts gives a good estimate of the entanglement entropy up to a multiplicative constant. In the vacuum, we can surmise this from the logarithmic scaling of entanglement entropy with interval size \cite{cardycalabrese}, which agrees with the number of cuts. This heuristic reasoning was verified numerically in \cite{mera, mera2, mera-expl}. More formal evidence was given in \cite{quotientmera}, which confirmed that the state on the exclusive causal cones of complementary semi-infinite lines has a thermal entaglement spectrum (see Sec.~\ref{sec:BHmera} for a more detailed summary). In light of this fact, the proportionality of entanglement entropy and the number of cuts is equivalent to the extensivity of thermal entropy. For any state built from networks with suitably generic tensors, the proportionality of entanglement entropy and the number of cuts will be established in upcoming work \cite{patrickmichael}, which draws on Page's theorem \cite{donpage}. 

In the present discussion of the MERA network, we treat the cut-counting prescription as an empirical fact. The reader should remember, however, that no fundamental principle protects this relation and it cannot be expected to hold in full generality. A case in point \cite{glenprivate} is the minimally updated MERA network \cite{minupdates}, which models a CFT in the presence of an impurity. In such circumstances, any connection between MERA and holography will involve the incremental entanglement entropy per bond instead of a na{\"\i}ve count of bonds.


\section{MERA and Kinematic Geometry}
\label{sec:therelation}

Kinematic space encodes the data about CFT subsystems in an elegant geometric way. CFT intervals are organized by location and scale in a Lorentzian space whose metric structure is supplied by conditional mutual information. An analogous representation of CFT subsystem data is given by the MERA network whose tensors are canonically associated with contiguous collections of UV sites. In this section, we outline a series of commonalities that motivate the identification of the two structures. In particular, we propose to view MERA as a discrete counterpart of kinematic space.

Our proposal to associate the MERA network with kinematic space runs contrary to a long-held belief that MERA ought to discretize a spatial slice of the bulk geometry. This idea, first put forward by Swingle \cite{briansessay, brianspaper}, gave the impetus to the prolific program of investigating tensor networks vis-{\`a}-vis holographic duality \cite{tadashisnetwork, hartmanmaldacena, xiaoliang, shocks, errorcorrectingnetwork, donspaper, patrickxiaoliang}, of which the present paper is a part. It is, thus, worthwhile to contrast our novel kinematic proposal with `the traditional view' of MERA as a discretized spatial geometry. In the discussion to follow, we comment on the conceptual drawbacks of a direct connection to the bulk, which are manifestly absent from the kinematic space perspective.\footnote{Our arguments are structural in character and differ fundamentally from the reasoning followed in \cite{critique}, which was based on counting degrees of freedom.}

\subsection{Partial order of MERA and kinematic causality}
The space of geodesics is a partially ordered set. This is an intrinsic property of kinematic space that follows from the containment relation of their boundary support---a property that is invariant under symmetry transformations. The signature of the kinematic metric is the geometric reflection of this structure.

The same applies to MERA: the tensors in the network are partially ordered with respect to their domains of influence. The locality of the tensor contractions, which is built into the skeleton of the network, makes each tensor capable of affecting only a subset of the spatial degrees of freedom. This immediately induces a hierarchy among them in that the regions affected by certain tensors are strictly enclosed within the domain of other tensors' influence. This property of MERA makes no reference to a UV cutoff. Moreover, the unitarity of the tensors promotes this ordering to a true notion of causality: Not only do individual tensors affect the state of well-defined spatial intervals, but also the state on given intervals is influenced only by specific network subregions. We can, therefore, draw light-like directions which restrict the propagation of information in the network.

We observe that the two notions of causality---network and kinematic---coincide. This structure is absent from the hyperbolic plane, all points in which are treated on equal footing. Only upon introducing a cutoff can points on $\mathbb{H}_2$ be partially ordered with reference to their distance from the boundary. We shall see that this structural difference has interesting consequences.

\subsubsection{Spacelike versus timelike paths}
\begin{figure}
	\centering
	\includegraphics[width=0.47\textwidth]{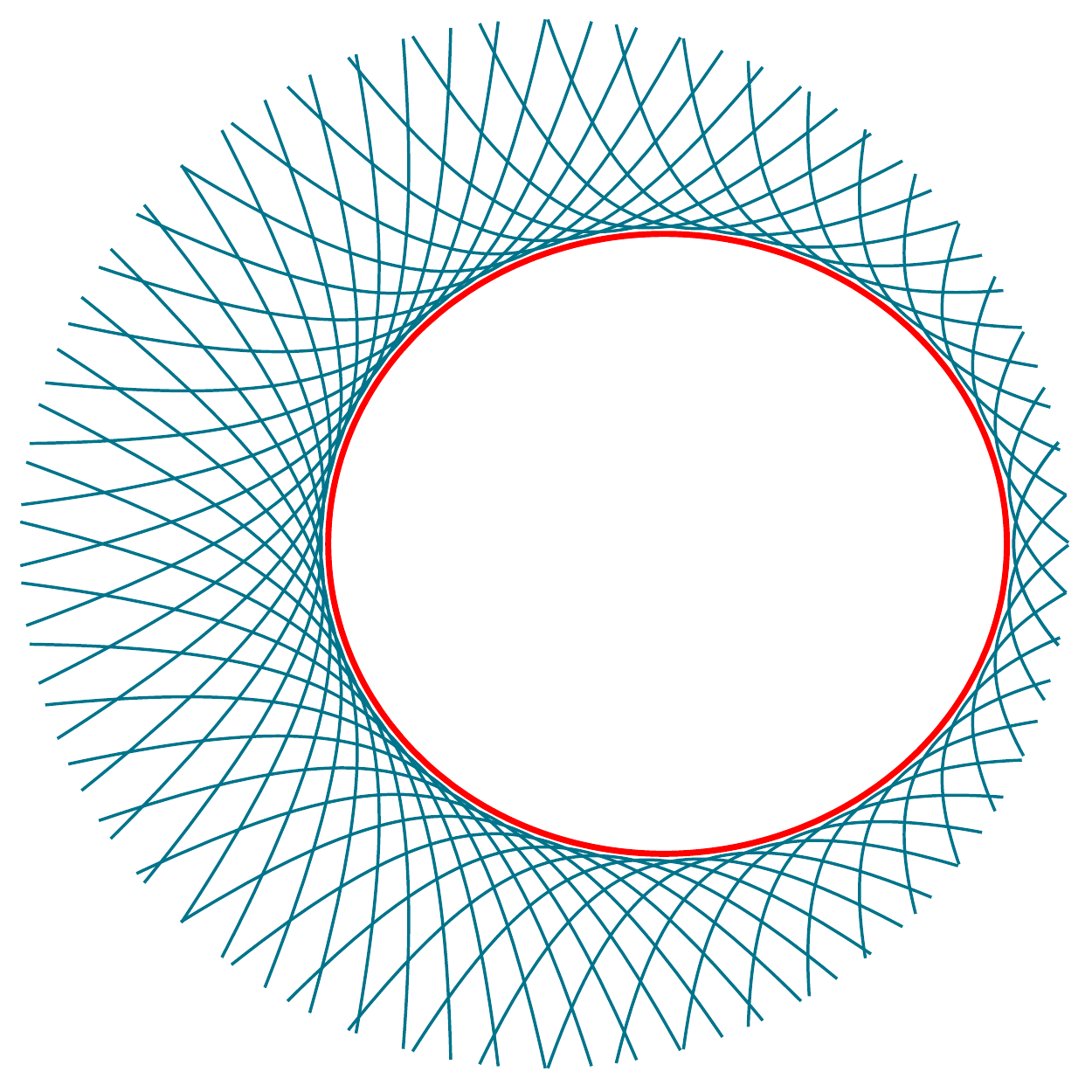}
	\hfill
	\includegraphics[width=0.47\textwidth]{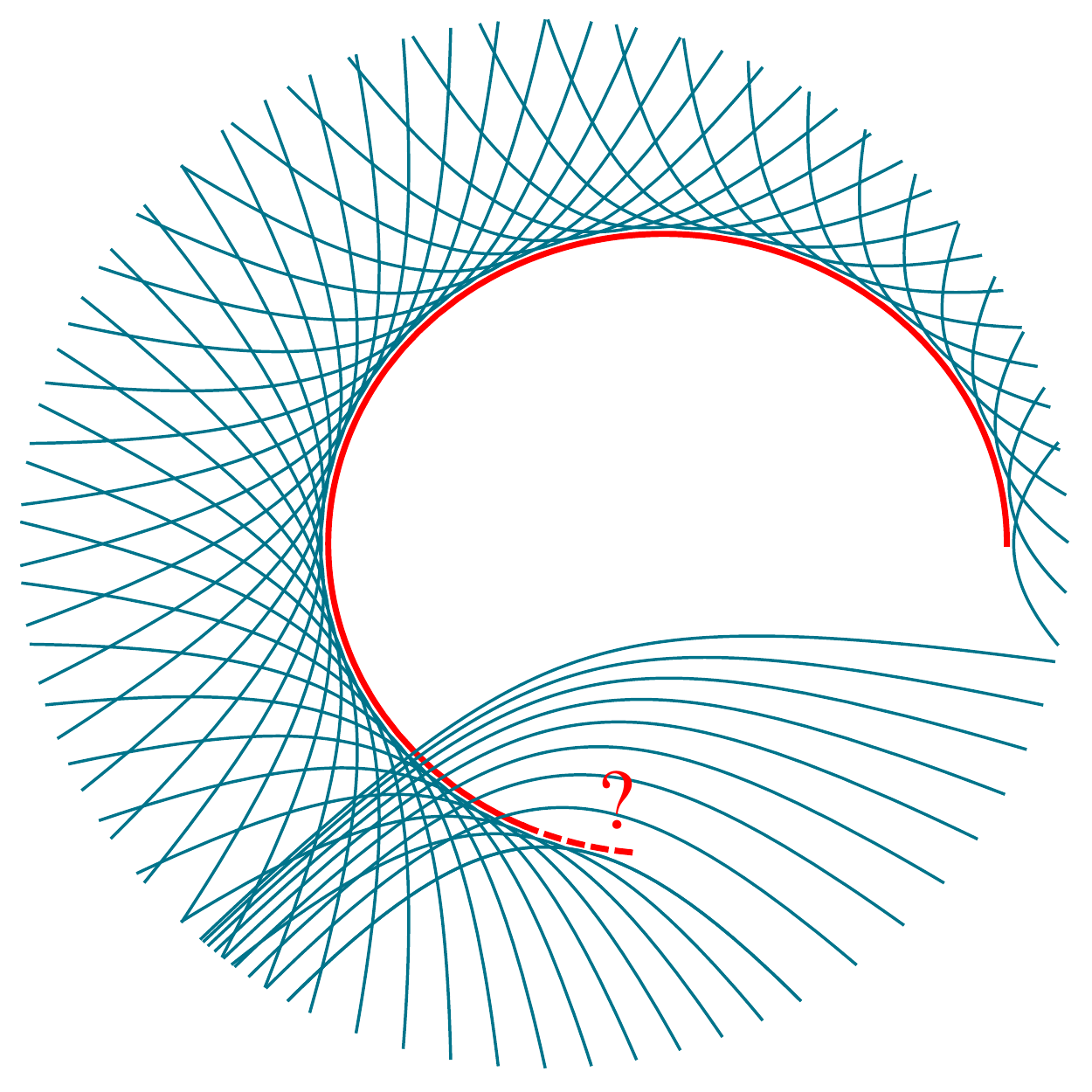}
	\caption{Spacelike (left) and timelike (right) curves in kinematic space as seen from the bulk point of view.}
	\label{fig:KScutoff}
\end{figure}

An immediate consequence of the Lorentzian signature of kinematic space is a qualitative distinction between kinematic paths that are \emph{spacelike}, \emph{null} or \emph{timelike}. This classification is robust under the action of symmetries and suggests that only certain types of curves, i.e. spacelike, can be used as good kinematic cutoff surfaces. The stipulation that cutoffs must not be timelike is evident in the holographic view of kinematic space. The reason is illustrated in Fig.~\ref{fig:KScutoff}. A spacelike kinematic trajectory selects a family of geodesics, which has a well-defined outer envelope in the bulk. This envelope acts as a (diffeomorphism invariant) cutoff surface in the spatial geometry. But whenever a trajectory in kinematic space becomes timelike, the bulk cutoff surface is no longer defined \cite{robfirst}.

Tensor networks, on the other hand, prepare wavefunctions on selected cutoff surfaces in the network. The causal structure of MERA, however, makes some cutoffs inadmissible: The cut on which the state is defined must be, like in kinematic space, piecewise spacelike or lightlike and never timelike \cite{quotientmera}. A MERA network ending on a locally timelike cut cannot be associated to a coarse-grained version of the vacuum wavefunction. The failure of MERA to prepare the wavefunction on a timelike cutoff surface is analogous to the failure of timelike-separated (in kinematic space) geodesics to form a curve in AdS. 

The Euclidean signature of a spatial AdS slice is compatible with any convex bulk cutoff surface \cite{AdSMERAtakayanagi}. The mixed signature of MERA is therefore in obvious tension with a direct AdS/MERA connection. This is best seen from the simple example of a cutoff that follows a radial geodesic in $\mathbb{H}_2$. Symmetry demands that such a cutoff surface, although geometrically admissible in the bulk, map to a vertical (and therefore timelike) cut in MERA. This conclusion can only be averted by introducing extra assumptions into a putative MERA/AdS correspondence.

\subsubsection{Representation of symmetries}
In studying MERA representations of CFT states, an important question concerns the action of conformal symmetry on the tensor network. Because the choice of a cutoff surface on which the state is defined breaks conformal invariance, the cutoff transforms under the conformal group. In two dimensions, conformal symmetry acts locally and can reset the cutoff to an arbitrary function of position.\footnote{Of course these statements hold up to the usual artifacts of discretization.} 

\paragraph{Local conformal transformations} The primary focus of the discussion in \cite{quotientmera} were local conformal transformations in MERA. A conformally transformed wavefunction was recognized as the state living on an inhomogeneous cut in the network. In this way, conformal maps in MERA are implemented by locally changing the cut on which the wavefunction is defined. Importantly, this operation does not affect the rest of the network away from the UV cut. In particular, the causal structure of the MERA network is fixed and independent of conformal transformations. To summarize, the action of the conformal group in MERA can take a uniform UV cutoff to some other, inhomogeneous cutoff, but without affecting the null directions. Consistency then requires that conformal maps take \emph{spacelike} cutoffs to other \emph{spacelike} cutoffs. 

Such a constraint is guaranteed when MERA is associated with the kinematic geometry. Interpreted in the bulk, however, this seems to impose an artificial restriction on the set of allowed (or MERA-representable) cutoff surfaces: they can never become approximately radial. Since conformal symmetry transforms radial and other bulk surfaces into one another, such a limitation would be a radical breaking of conformal symmetry.

\paragraph{Inhomogeneity of a causal cut} Causal cuts in MERA are not homogeneous. Their lightlike segments are uniform, but the top of a causal cut where left-going and right-going cuts meet is distinct from the rest. The non-uniform shape of a causal cut in MERA is readily understood in the kinematic interpretation. The top corresponds to the geodesic $g$ supported on the base of the chosen lightcone while other points on a causal cut correspond to narrower geodesics that share one endpoint with $g$ and are otherwise contained within it (compare e.g. Fig.~1 in \cite{protocol}). The insensitivity of both the MERA and kinematic partial order to the UV-cutoff ensures that this point will remain special under local conformal maps.

The AdS isometries, on the other hand, map different points on the same geodesic to one another. In other words, geodesics are homogeneous, a fact that forbids special points. When identifying the causal cut with an AdS geodesic -as the direct AdS/MERA connection suggests- one might try to assuage this discrepancy by declaring that the special point on a cut in MERA corresponds to some select point on a bulk geodesic, chosen according to some prescription. Any such prescription, however, must refer to a UV cutoff; in the absence of a UV cutoff there is no reference with respect to which a special point may be chosen. Because conformal symmetry acts on the cutoff, it must also affect the choice of a preferred point on a geodesic. Yet in MERA, the top of a causal cut is fixed, its location blind to any changes in the cutoff. This reveals that the conformally invariant notion of causality in MERA disfavors a na{\"\i}ve partial ordering of the hyperbolic plane induced by a UV cutoff. But it is in full agreement with the  causal structure of kinematic space, which is likewise conformally invariant.

\subsection{Localization of information}

\subsubsection{Crofton form and volumes in MERA}
\label{meravolume}

In Sec.~\ref{kincmi}, we observed that the notion of volume of kinematic space (eq.~\ref{kinvolume}) hails from information theory: it is the conditional mutual information (\ref{defcmi}) of three contiguous intervals. Let us inspect the same quantity in MERA.

\begin{figure}
	\centering
	\includegraphics[width=1.\textwidth]{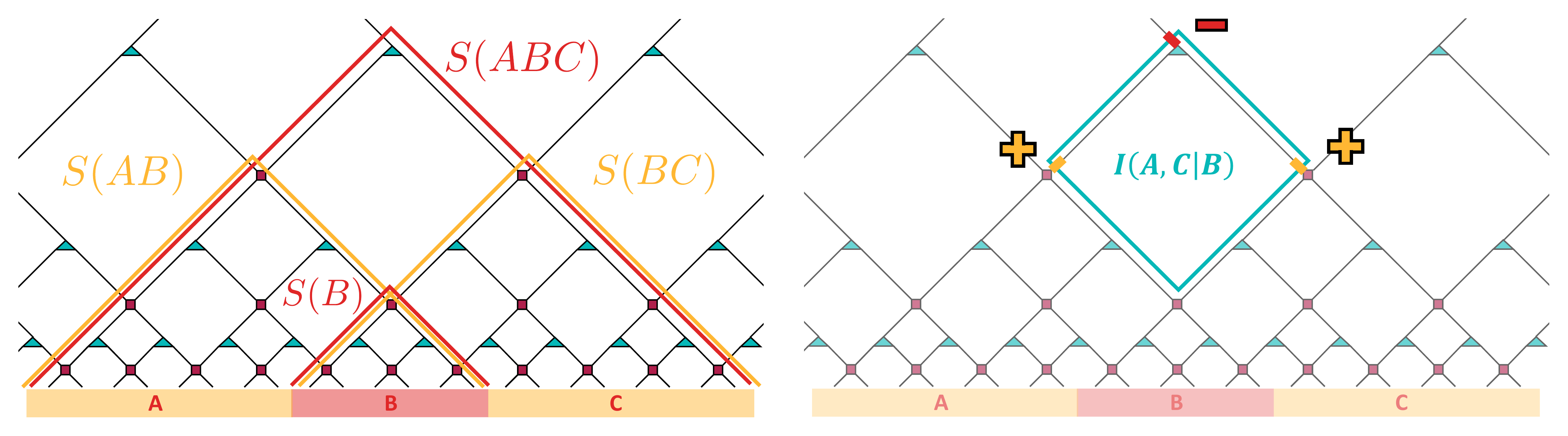}
	\caption{Conditional Mutual Information in MERA. Most cuts that contribute to the computation of $I(A,C|B)$ cancel in the alternating sum. The net contribution to the conditional mutual information arises from a localized region of the network.}
	\label{fig:meraCMI}
\end{figure}

\paragraph{Conditional mutual information localizes in MERA} When we apply the cut-counting prescription reviewed in Sec.~\ref{cutcount} to
\begin{equation}
I(A, C|B) = S(AB) + S(BC) - S(ABC) - S(B)\,,
\tag{\ref{defcmi}}
\end{equation}
we obtain Fig.~\ref{fig:meraCMI}. The cuts associated with the positive terms in (\ref{defcmi}) are in large part the same as the cuts for the negative terms, leading to cancellations. The net result comes from a localized part of the network, whose boundaries are lightlike. In other words, the conditional mutual information of neighboring intervals localizes in a causal diamond. For intervals with endpoints at
\begin{equation}
A = (u - \Delta u, u)
\qquad {\rm and} \qquad 
B = (u, v) 
\qquad {\rm and} \qquad  
C = (v, v + \Delta v)\,,
\label{3INT}
\end{equation}
the relevant causal diamond resides between $u$ and $u-\Delta u$ in the left-moving coordinate and between $v$ and $v + \Delta v$ for the right-moving one. 

\begin{figure}
	\centering
	\includegraphics[width=.7\textwidth]{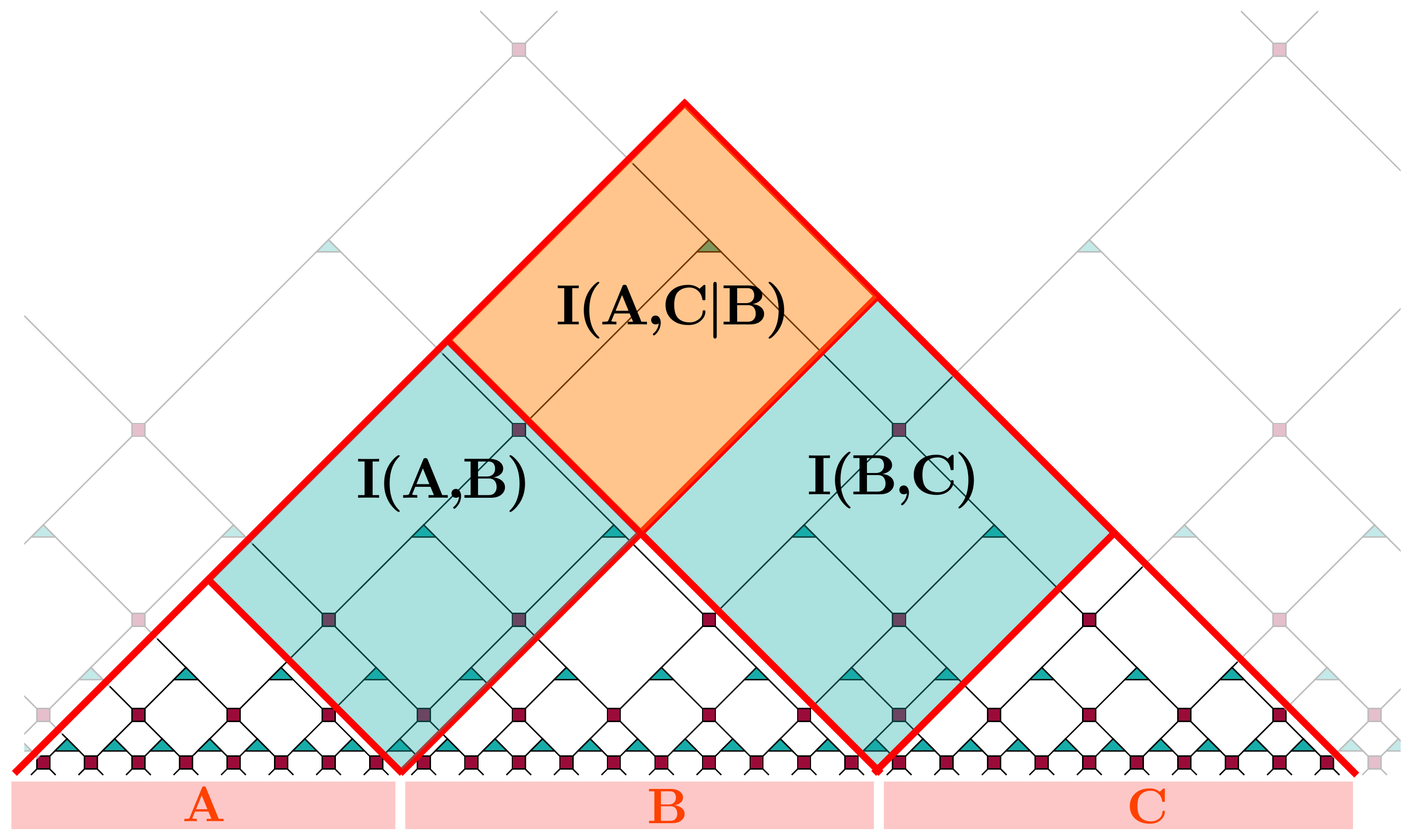}
	\caption{Localization of mutual information in MERA. We indicate the local regions of the network that control the computation of the mutual information of two neighboring intervals, and the conditional mutual information of three neighboring intervals.}
	\label{fig:meratiles}
\end{figure}

In the end, the entire MERA network is a tilted chessboard of causal diamonds, each of which computes some conditional mutual information. The grid of lightlike coordinates demarcates conditional mutual informations of different triples of intervals. Fig.~\ref{fig:meratiles}, which displays these facts, is a faithful copy of Fig.~\ref{fig:KSmutual}, which highlights the analogous characteristics of kinematic space.

\paragraph{What does conditional mutual information count?} Figs.~\ref{fig:meraCMI}~and~\ref{fig:meratiles} give a crisp answer: conditional mutual information counts how many isometries live in the appropriate causal diamond. Eq.~(\ref{defcmi}) asks for the net reduction in the number of lines passing through the causal diamond as we go from the bottom up. The only way we can register a net loss of lines is if a line is soaked up by an isometry. Indeed, every isometry accounts for precisely one line, which enters the diamond from the bottom but does not emerge at the top. Counting the decrease in the number of lines is equivalent to counting isometries.

\paragraph{Conditional mutual information as volume} 
We propose to adopt conditional mutual information as a definition of volume in MERA. The two facts highlighted above guarantee that this is a reasonable proposal:
conditional mutual information localizes in MERA and counts a crisply defined object---the isometries contained in a causal diamond. In other words, we observe that for $A, B, C$ defined in eq.~(\ref{3INT}):
\begin{equation}
\mathcal{D}\textrm{(isometries)} = I(A, C | B)  \,.
\label{disometries}
\end{equation}
We declare this quantity a discrete volume form, in analogy to eq.~(\ref{kinvolume}) in kinematic space.

In the upcoming second part of this work \cite{compression}, where we discuss our more general compression networks, we will appreciate better the rationale for working with eq.~(\ref{disometries}). The volume of a causal diamond computed by (\ref{disometries}) evaluates the amount by which the tensors in the diamond compress the state living on its past edges. This is how eq.~(\ref{disometries}) should be viewed in applications beyond the standard MERA. In the special case of the vacuum MERA, this `density of compression' is directly proportional to a na{\"\i}ve count of isometries. We give a short summary of the compression networks in Sec.~\ref{compressionsummary}, referring to \cite{compression} for details.

It is worth noting that in the traditional holographic view of MERA the connection between conditional mutual information and localized volumes of the network is puzzling. If we represent the terms in eq.~(\ref{defcmi}) by geodesics in the bulk, no such localization occurs. Instead, the calculation involves an extended region in the spatial geometry, which reaches all the way to the asymptotic boundary. When $A$ and $C$ are taken to be small as in eq.~(\ref{3INT}), the bulk region associated with $I(A, C | B)$ becomes a fattened geodesic subtending $B$. In MERA this limit shrinks the relevant causal diamond to a small number of tensors. This again motivates relating small regions in MERA to bulk geodesics.

\paragraph{A metric for MERA} In Sec.~\ref{kincausality}, we assembled the kinematic metric (\ref{kinmetric}) from two ingredients: the causal structure (eq.~\ref{nullcoords}) and the volume form (\ref{kinvolume}). In Sec.~\ref{MERAcausality} we recognized that MERA has an identical causal structure. Now eq.~(\ref{disometries}) gives us a notion of volume, which is a direct analogue of eq.~(\ref{kinvolume}). These reasons justify conceptualizing MERA as a discrete version of kinematic space. More explicitly, we may write down a discrete tensor network metric
\begin{equation}
ds^2_{\rm T.N.} = I(\Delta u, \Delta v | B) 
\,\,\stackrel{\rm MERA}{\xrightarrow{\hspace*{12mm}}}\,\, 
(\# {\rm isometries}) \, \Delta u \, \Delta v
\label{ds2tn}
\end{equation} 
which in the case of the familiar MERA simply counts isometries in causal diamonds. This metric is the obvious counterpart to eq.~(\ref{kinmetric}) in kinematic space.

\paragraph{Differential entropy and cut-counting in MERA} One attractive feature of kinematic space is that volumes in it reproduce the differential entropy formula \cite{holeography}; see eq.~(\ref{sdiff}). Metric~(\ref{ds2tn}) ought to give rise to a similar relation in MERA.

Indeed, any spacelike cut across MERA defines a (possibly non-uniform) UV cutoff and a coarse-grained Hilbert space; see Sec.~\ref{merastructure}. The logarithm of the dimension of that Hilbert space is proportional to the number of indices living on the cut. Because every line ends on some isometry in the UV part of the network, the logarithm of the dimension of the Hilbert space defined by a cut is equal to the volume of MERA living above that cut, counted according to eq.~(\ref{disometries}). The equality between the `volume' of a subregion of MERA and the number of lines on its boundary follows from a discrete version of Stokes' theorem.

This argument is an exact analogue of the reasoning articulated below eq.~(\ref{sdiff}). Thus, computing the size of a coarse-grained Hilbert space by counting indices on its defining cut is the MERA version of the differential entropy formula. As a special case, this recovers the cut-counting prescription for entanglement entropy, which we revisit in Sec.~\ref{entanglementMERA}. More generally, counting cuts assigns an entropic quantity to any (possibly non-uniform) spacelike UV cutoff, which in the bulk is represented by a collection of tangent geodesics (see Fig.~\ref{fig:KScutoff}).

\subsubsection{Entanglement entropy}
\label{entanglementMERA}

The feature of MERA that makes it especially relevant for holography is the way it geometrizes entanglement entropies. In Sec.~\ref{cutcount} we reviewed the cut-counting prescription in MERA: estimating the entanglement entropy of an interval by counting the lines which cross the causal cut. This special property of the optimized network was used in Sec. \ref{meravolume} to place a physical metric on MERA (eq. \ref{ds2tn}) and recognize it as a faithful representation of the kinematic geometry.


\begin{figure}
	\centering
	\includegraphics[width=0.57\textwidth]{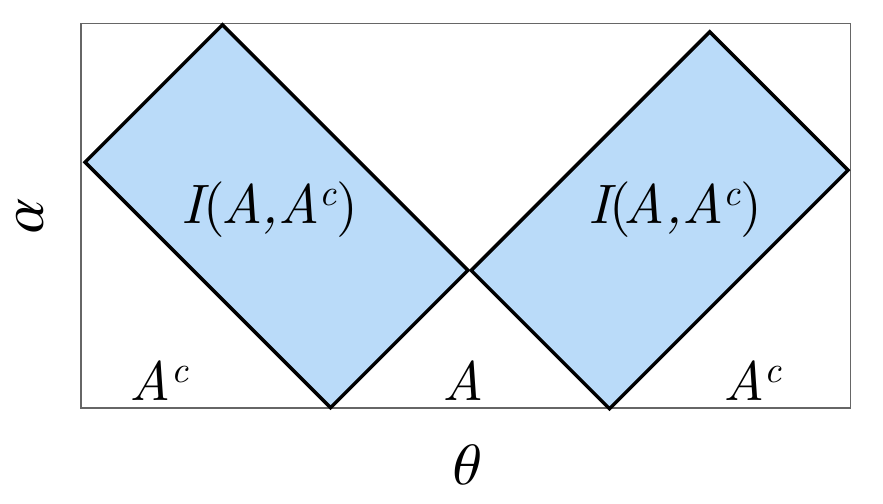}
	\caption{The entanglement entropy of $A$ is given by half of the total kinematic volume of the `causal wings' depicted in the figure.}
	\label{fig:KSentropy}
\end{figure}

If MERA discretizes the kinematic space, however,  the causal cut in MERA becomes a discrete version of the kinematic causal cone. The integral geometric computation of entanglement entropies then ought to be consistent with the cut-counting prescription along this causal cut. Recall that in a pure state, the entanglement entropy of an interval $\mathcal{I}$ is half the mutual information of $\mathcal{I}$ and its complement, $\mathcal{I}^c$:
\begin{equation}
S(\mathcal{I}) = \frac{1}{2}\, I(\mathcal{I}, \mathcal{I}^c)
\label{eecmi}
\end{equation}
As we saw in Sec.~\ref{kincmi}, the mutual information of two adjacent intervals can be read off from kinematic space as the volume of a causal diamond, which includes the common endpoint of both intervals. In the case at hand, we actually have two causal diamonds, because $\mathcal{I}$ and $\mathcal{I}^c$ have two endpoints in common (see 
Fig.~\ref{fig:KSentropy}.) We may use eq.~(\ref{sdiff}) and convert the volume of the two causal diamonds to a differential entropy, that is a one-dimensional integral over the causal cut:
\begin{equation}
S(u,v) = 
\frac{1}{2} \int_u^v d\tilde{v}\,\,\frac{\partial S(u,\tilde{v})}{\partial \tilde{v}} +
\frac{1}{2} \int_v^u d\tilde{u}\,\,\frac{\partial S(\tilde{u},v)}{\partial \tilde{u}}
\label{eece}
\end{equation}
The two terms in this formula come from the two causal diamonds in Fig.~\ref{fig:KSentropy}. In the context of MERA, their integrands become densities of lines that cross the causal cut (Sec. \ref{meravolume}). This is precisely what the cut-counting prescription mandates.

\begin{figure}
	\centering
	\includegraphics[width=0.57\textwidth]{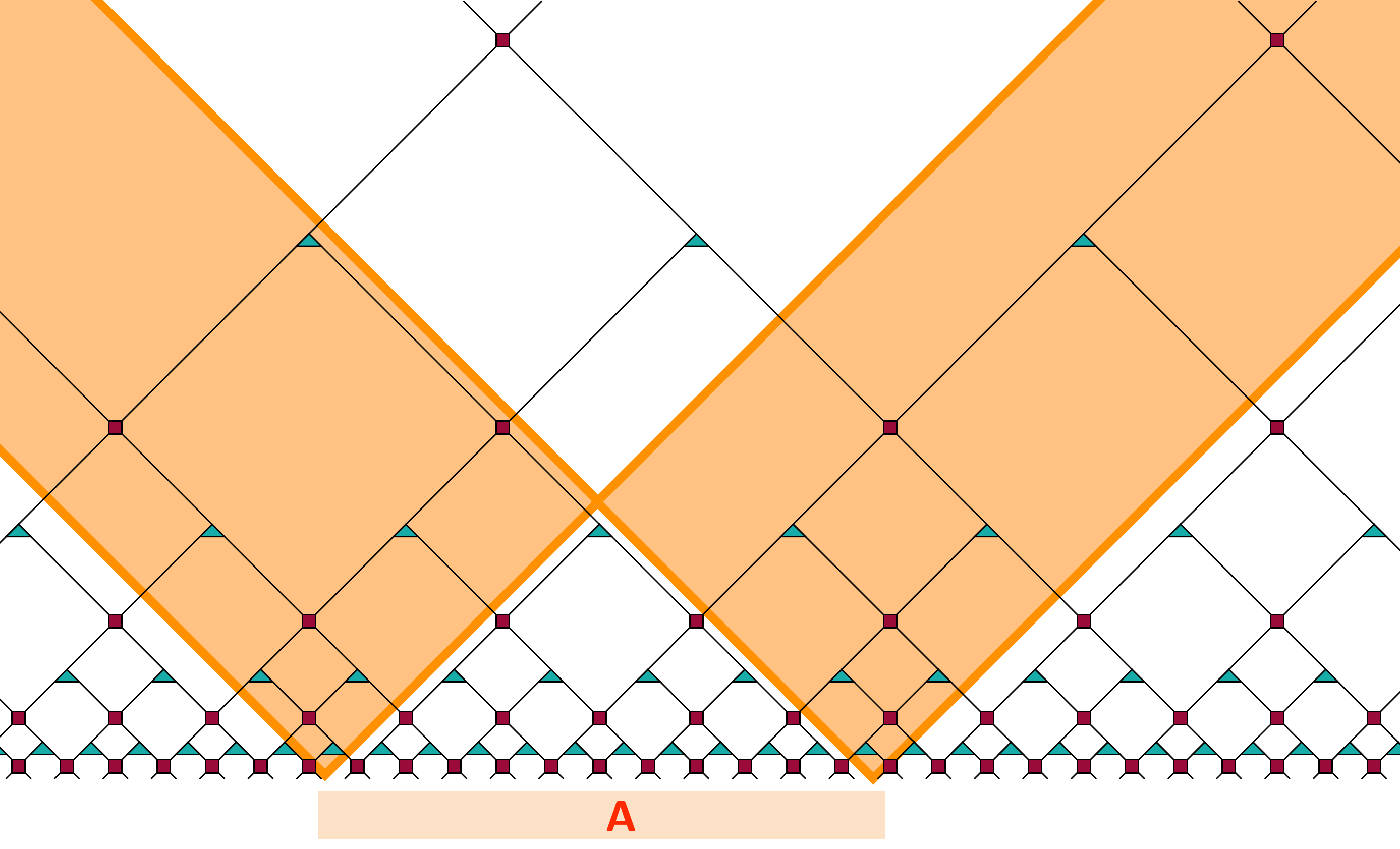}
	\hfill
	\includegraphics[width=0.4\textwidth]{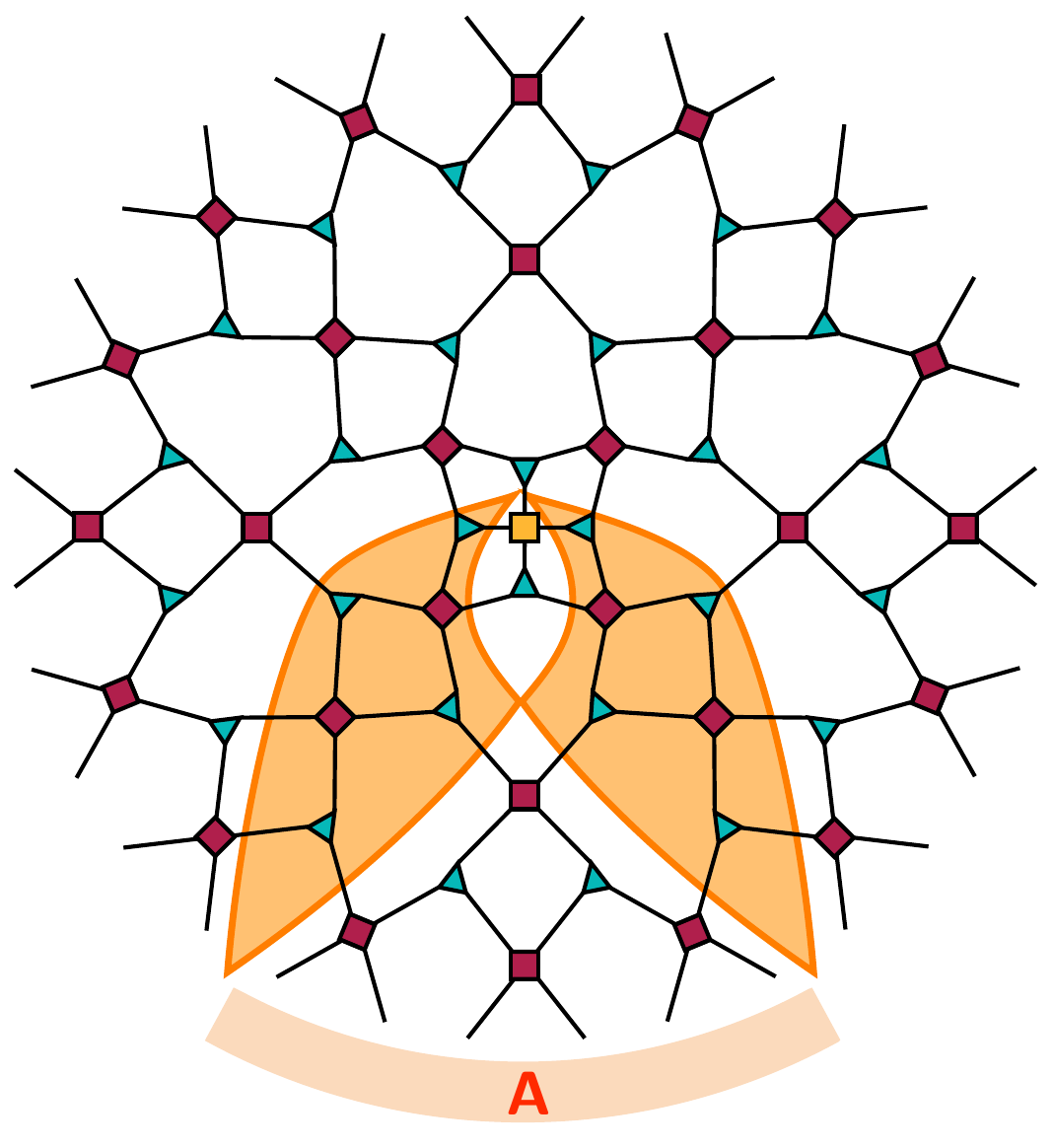}
	\caption{The entanglement entropy of a CFT interval $A$ is computed by the number of network isometries responsible for correlating $A$ with its complement. This is consistent with the integral geometric computation of the corresponding geodesic length (Fig.~\ref{fig:KSentropy}). An application of the discrete Stokes' theorem recovers the cut-counting prescription of Sec.~\ref{cutcount}. The figure shows the network volume relevant for the EE on the line (LEFT) and the circle (RIGHT). It contains all tensors in the inclusive causal cones of the two endpoints of $A$, excluding their intersection. This network region prepares the entanglement spectrum between $A$ and $A^c$.} 
	\label{fig:meraEE}
\end{figure}

To see this more clearly, consult Fig.~\ref{fig:meraEE}, which is the MERA analogue of Fig.~\ref{fig:KSentropy}. Taking advantage of the localization of mutual information in MERA illustrated in Fig.~\ref{fig:meratiles}, we know that the entanglement entropy of $A$ is given by the number of isometries living in the highlighted part of the network. But the same count yields the number of lines crossing the causal cut of $A$. In effect, we are learning that the cut-counting prescription of Sec.~\ref{cutcount} secretly enumerates 
%
the isometries responsible for correlating the interval with its complement. This is in direct analogy with the way kinematic space encodes the length of a Ryu-Takayanagi geodesic as the `number of geodesics' connecting the boundary interval with its complement. The counting of geodesics is done with the Crofton measure, which is a geometric counterpart of the density of isometries in MERA.

In contrast, if we place the network directly on the time-slice of AdS, the relation between volumes of `causal wings' and entanglement entropies appears mysterious. It seems to imply that a special region in the bulk---which lacks an independent  motivation in the AdS/CFT correspondence---quantifies the correlation between a given interval and its complement in terms of its volume; see Fig.~\ref{fig:meraEE}. Insisting on an AdS/MERA correspondence appears to add another peculiar property to its putative dictionary, a peculiarity that is readily resolved by the kinematic proposal.

\subsection{MERA as renormalization}
\label{merarg}	

\subsubsection{Coarse-graining with MERA}
\label{coarsemera}
As we reviewed in Sec.~\ref{merastructure}, MERA provides a graphical representation of renormalization in real space. The vertical direction corresponds to scale in the field theory. Cutting MERA on different levels defines states, which are related to one another by coarse-graining or fine-graining. As we go higher up in MERA, the successively coarse-grained states live in Hilbert spaces of exponentially decreasing sizes (entropies).

The same features are observed in kinematic space; its identification with the space of CFT intervals makes it a natural domain for real space cutoffs. The two coordinates of kinematic space, $\theta$ and $\alpha$ (see eq.~\ref{nullcoords}) also correspond to location and scale. The role of $\alpha$ as setting a scale is evident from its definition as the half-width of a field theory interval. Cutting off kinematic space at $\alpha = \alpha_*$ imposes a real space cutoff---it amounts to declaring $2\alpha_*$ to be the smallest resolution in the field theory. The spatial size of a cutoff surface in kinematic space also varies exponentially with the cutoff; in the vacuum on a circle, metric (\ref{kinmetric}) expressed in terms of $\theta$ and $\tilde\rho = -\log (\csc\alpha + \cot\alpha)$ is:
\begin{equation}
ds_{\rm kin}^2 = \frac{c}{3}\, (-d\tilde\rho^2 + \cosh^2\tilde\rho \, d\theta^2)\,.
\label{metrickin}
\end{equation}

Holographically, every real space cutoff defined by a curve in kinematic space selects a set of bulk geodesics. These in turn identify a bulk cutoff surface by their outer envelope as we illustrated in Fig.~\ref{fig:KScutoff}. This proposal for the holographic cutoff has the appealing feature that it is manifestly diffeomorphism invariant, because it is implemented on bulk geodesics that make no reference to AdS coordinate systems. Interestingly, the kinematic cutoff can be further promoted to a precise coarse-graining prescription for CFT operators, which exploits the structure of the operator product expansion (OPE). A detailed discussion of this point will be presented in \cite{kinematicoperators}, where we formulate 1+1-dimensional CFTs in the language of kinematic space and derive the connection to the effective field theory in the bulk from first AdS/CFT principles.

On a spatial slice of the bulk geometry, the radial direction $\rho$ is also dual to changes of scale in the CFT \cite{uvir1, uvir2, uvir3}. Regulating large scale divergences on the gravity side with a radial cutoff $\rho = \rho_*$ is dual to selecting an ultraviolet cutoff in the CFT. 
When we push the radial cutoff $\rho_*$ to infinity, the area of the cutoff surface grows exponentially. This is captured by the spatial metric:
\begin{equation}
ds^2 = L^2 (d\rho^2 + \sinh^2\rho\,  d\theta^2)
\label{metricspatial}
\end{equation}
The interpretation of MERA as a real space RG transformation, however, can be leveraged to distinguish between the two types of geometric coarse-graining suggested above. As we explain in the next section, the constraints that causality imposes on the RG operation of MERA act in favor of the kinematic proposal.


\subsubsection{Real space RG and causal cuts}
\label{rgcausalcuts}

Consider two points on the 1-D boundary slice where the CFT state lives. Any such choice splits the CFT into two regions: an interval $A$ and its complement $A^c$. 
In a pure state such as the vacuum the entanglement entropies $S(A)$ and $S(A^c)$ are equal. This fact is nicely captured by the Ryu-Takayangi proposal: a unique minimal geodesic homologous to both $A$ and $A^c$ joins the two boundary points.

\begin{figure}
	\centering
	\includegraphics[width=0.53\textwidth]{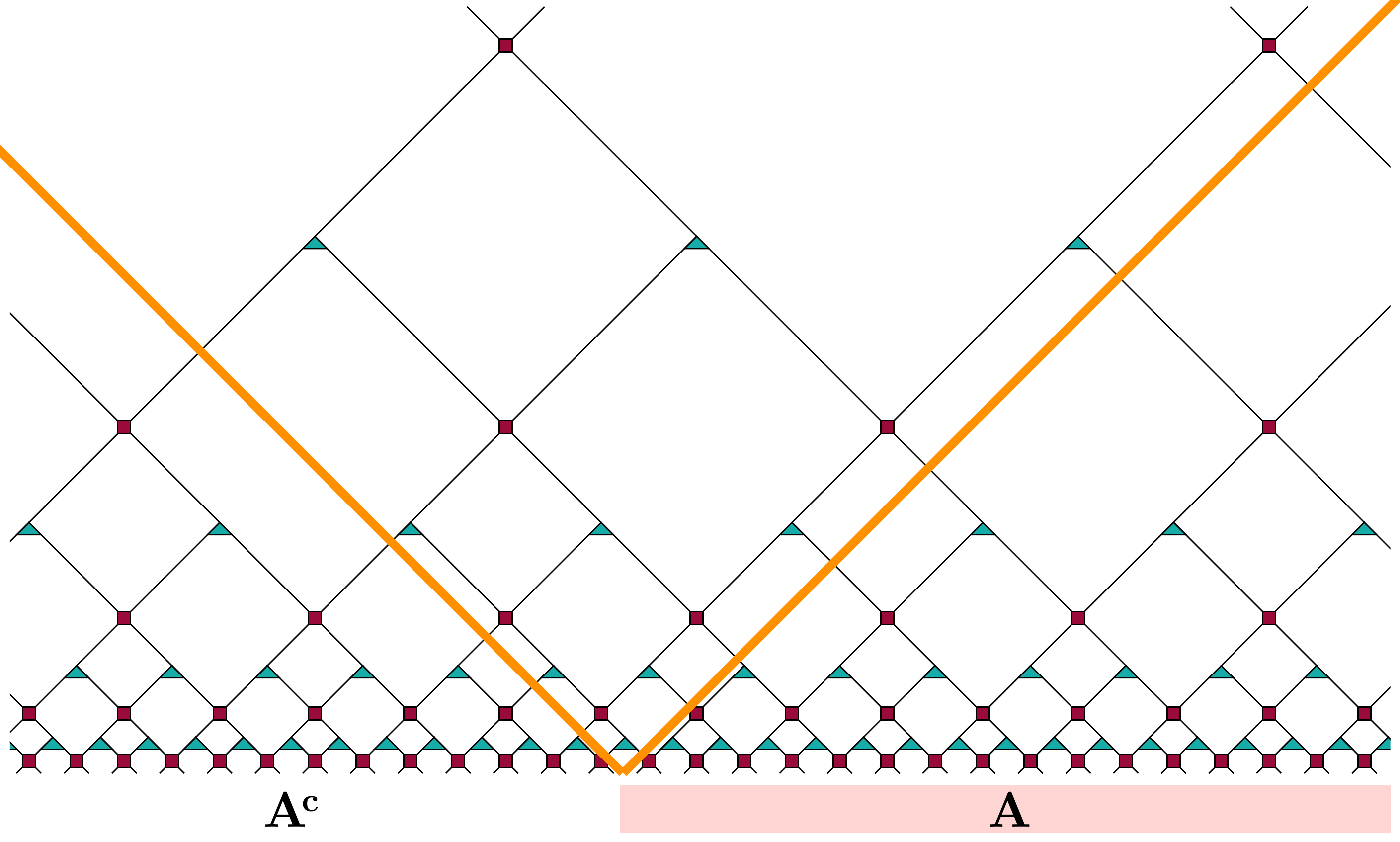}
	\hfill
	\includegraphics[width=0.4\textwidth]{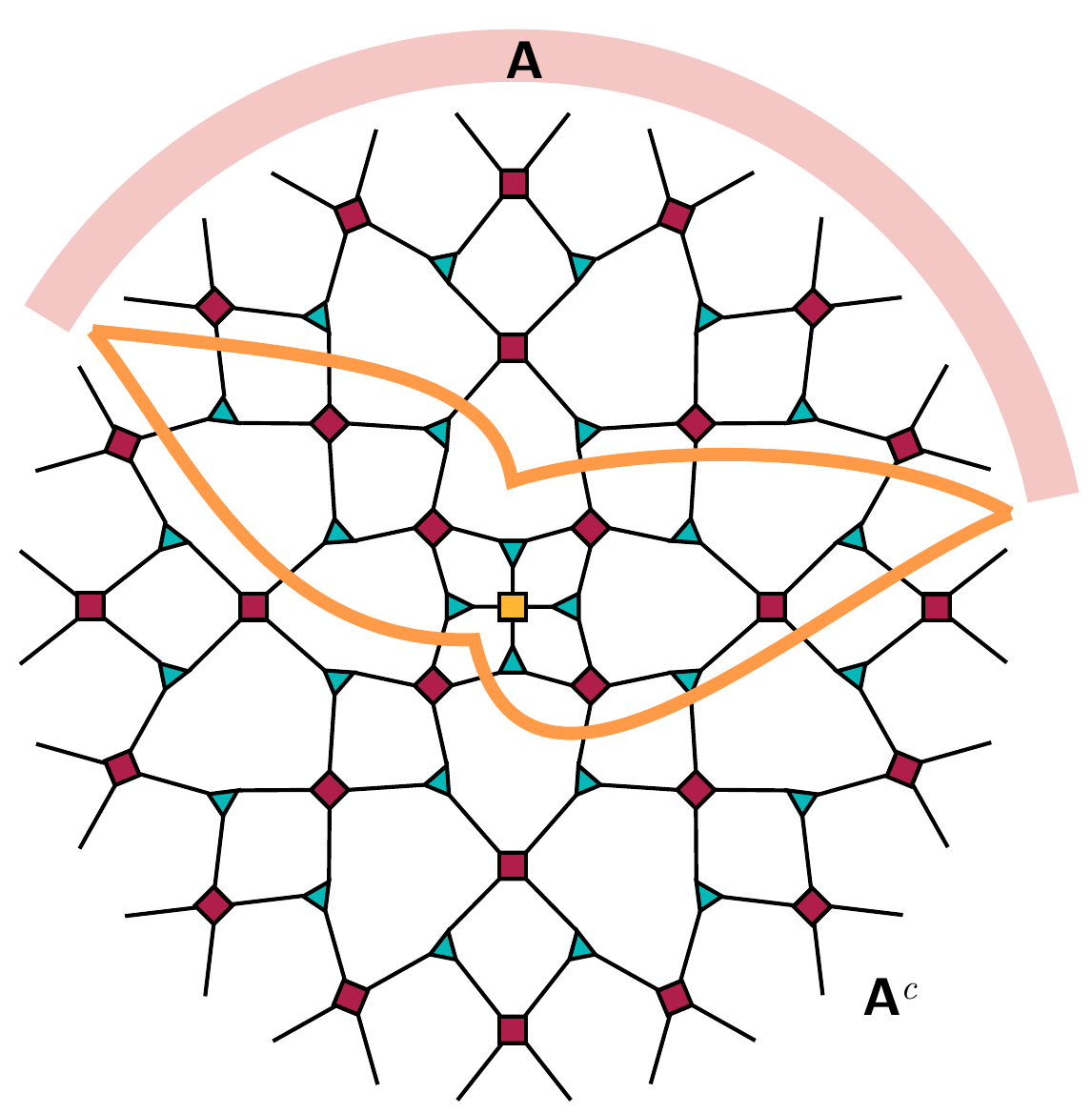}
	\caption{Causal cuts for a region $A$ and its complement $A^c$. For both MERA on the line and circle, the causal cuts for $A$ and its complement $A^c$ are distinct.}
	\label{fig:radial-cut}
	\end{figure}

In MERA, for any selection of two spatial points there are two distinct causal cuts in the network, which bound the exclusive causal cones of $A$ and $A^c$, respectively (see Fig. \ref{fig:radial-cut}). The two cuts typically do not cross the same number of links; only the minimal one has the correct count of links to match the entanglement entropy. Nevertheless, both lightcones are physically meaningful. In the view of MERA as a real space RG transformation, every local application of disentanglers and isometries performs a local coarse-graining of the wavefunction. Such coarse-grainings can be understood as a change of basis, but only if the cutoff surface is piecewise spacelike or null at every RG step. In this way, we obtain an upper bound for the allowed local coarse-graining of an interval.
The two causal cuts encode the maximally coarse-grained state of $A$ and $A^c$, respectively.

To appreciate the importance of the two distinct MERA lightcones in a different way, observe that the tensors contained in the exclusive causal cone of an interval build local correlations among the interval's lattice sites. 
This amounts to selecting a local basis for representing the state and leaves the entanglement spectrum unaffected. In other words, the entanglement spectrum of $A$ is solely prepared in the region between the two causal cones. 

This feature of MERA is naturally included in the kinematic proposal: the two complementary intervals possess distinct causal cones, each of which bounds the set of bulk geodesics anchored on the respective boundary region. 
There exists, moreover, a finite `volume' of geodesics that connect the two intervals, a fact reflected by the separation of the two lightcones in kinematic space. 
When approached from the traditional AdS/MERA perspective, however, no meaningful geometric counterpart exists for the non-minimal causal cut. 
This contradicts the equal treatment of the two cuts in the network and seems to select a peculiar, IR-probing curve associated to the coarse-grained state of the larger interval. 
By the AdS/MERA interpretation, that coarse-grained state should have been instead associated to the minimal geodesic. 

\section{MERA for boundary gravitons and two-sided black holes} 
\label{sec:BHmera}
\subsection{Boundary gravitons}

Thus far, we have argued for identifying the vacuum MERA with the kinematic space of a time slice of pure AdS$_3$. This conclusion automatically extends to conformal descendants of the vacuum---states related to the vacuum by a local conformal transformation. In MERA, wavefunctions of such states can be read off from inhomogeneous UV cuts \cite{quotientmera}. In particular, going from the vacuum to a descendant does not change local properties of the network. On the bulk side, descendant states are represented by so-called boundary gravitons \cite{banados}. They are locally AdS$_3$ geometries, which differ from global AdS$_3$ by large diffeomorphisms. Importantly, a large diffeomorphism changes lengths of geodesics by two additive pieces, which carry no joint dependence on the two endpoints \cite{mandal, joan}:
\begin{equation}
S(u,v) \to S(u,v) + \Delta \mu(u) + \Delta \mu(v)
\end{equation}
This change leaves kinematic volumes (\ref{kinvolume}) invariant. We reach the same conclusion by noting that a boundary conformal transformation that preserves a time slice of the CFT maps $x \to \tilde{x} = f(x)$. Applying this transformation to $u$ and $v$ in the kinematic metric (\ref{kinmetric}) gives:
\begin{equation}
ds_{\rm kin}^2= 
\frac{\partial^2 S(u,v)}{\partial u \,\partial v}\,
du\,dv 
=
\frac{\partial^2 S(\tilde{u},\tilde{v})}{\partial \tilde{u} \,\partial \tilde{v}}\,
d\tilde{u}\,d\tilde{v} 
\end{equation}
This illustrates that the kinematic space defined in eq.~(\ref{kinmetric}) is invariant under all conformal transformations which preserve a time slice of the CFT. The only dependence on the conformal frame is introduced by the UV cutoff.

\subsection{The thermofield double state and the two-sided BTZ black hole}

\begin{figure}
	\centering
	\includegraphics[width=.7\textwidth]{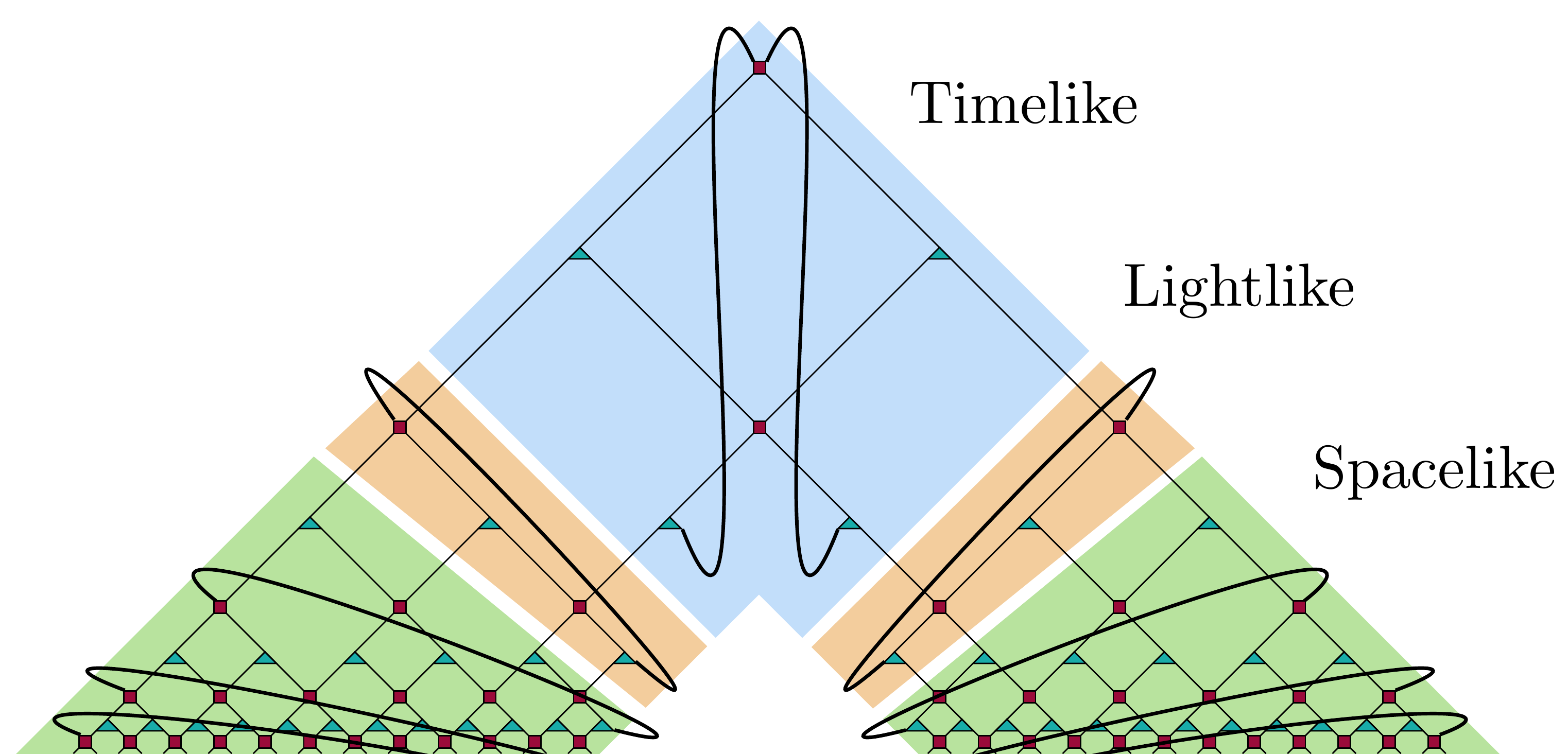}
	\caption{The quotient of the optimized vacuum MERA, which prepares the thermofield double state. In different parts of the network the identifications act in a timelike, lightlike and spacelike manner, respectively. The network displayed here identifies lines that are $k=2$ layers apart, so $s = 2^k = 2^2$ in eq.~(\ref{ltobeta}). The figure is reproduced with permission from \cite{quotientmera}.}
	\label{quotientTN}
\end{figure}

\paragraph{The thermofield double state in the CFT and in MERA}
For a non-trivial application of local conformal transformations, consider a map that acts not on the full line $\mathbb{R}$, but the line minus a point, $\mathbb{R}\setminus\{0\}$. The two semi-infinite lines on either side of the excluded point can each be mapped to an infinite line by the logarithmic map:
\begin{equation}
x \to (\beta / 2 \pi) \log |x| 
\label{logmap}
\end{equation}
In this way, we view the vacuum on $\mathbb{R}$ as an entangled state on $\mathbb{R} \times \mathbb{R}$. Famously, in a conformal field theory this entangled state is the thermofield double state. One may further quotient the two $\mathbb{R}$s by a discrete translation $\log x \sim \log x + \log s$ to obtain the thermofield double state on $S^1 \times S^1$. The circumference of the $S^1$ sets a natural scale, in which to express the otherwise dimensionless inverse temperature:
\begin{equation}
(\beta / 2\pi) \log s \equiv 2 \pi L
\label{defs}
\end{equation}

In Ref.~\cite{quotientmera} our collaborators and we performed these operations in the optimized MERA network. The conformal transformation (\ref{logmap}) was enacted by cutting the network along the two null rays emanating from $x=0$. After the cut, the quotient identifies identical pieces of the causal cone of the origin. Alternatively, we can apply the quotient prior to the conformal map (\ref{logmap}). This produces the entire network shown in Fig.~\ref{quotientTN}, including the regions living below the null rays. In this view, the two semi-infinite lines are modded out by a discrete scaling transformation $x \sim s\, x$.

\paragraph{The BTZ black hole as a quotient of AdS$_3$}
The field theory operations outlined above reflect a famous fact in 3-d gravity: that the two-sided BTZ black hole is a quotient of pure anti-de Sitter space \cite{btz}. Consider the Poincar{\'e}-AdS$_3$ metric restricted to $t=0$:
\begin{equation}
ds^2 = \frac{dx^2 + dz^2}{z^2}
\end{equation}
In order to quotient $\mathbb{R}_-$ and $\mathbb{R}_+$ by a discrete scale transformation, select a family of geodesics centered at $x = 0$ whose radii are related by powers of $s$:
\begin{equation}
x^2 + z^2 = s^{2n} r^2 \qquad {\rm where}~n \in \mathbb{Z}
\label{identgeos}
\end{equation}
Identifying these geodesics with one another produces a topological cylinder, which is the static slice of the two-sided BTZ geometry; see Fig.~\ref{BTZidents}. This identification can be canonically extended away from the time slice to produce the full, 2+1-dimensional BTZ space-time with two asymptotic regions. The inverse temperature of the black hole in units of the AdS$_3$ curvature scale is given by eq.~(\ref{defs}):
\begin{equation}
\beta / L = 4\pi^2 / \log s
\label{ltobeta}
\end{equation}

In the limit $n \to -\infty$, the geodesics in Fig.~\ref{BTZidents} zoom on a point on the boundary at $x = 0$. 
This location, which is a fixed point of the quotiented discrete scale transformation, separates the two semi-infinite lines into which the $x$-axis decomposes under map (\ref{logmap}). After the quotient, every fundamental domain in $x > 0$ represents a copy of one asymptotic boundary while fundamental domains in $x < 0$ are copies of the other asymptotic boundary. In the bulk, the line $x = 0$ is also meaningful. It connects points of closest approach of identified geodesics and, therefore, comprises images of the bifurcation horizon. Many good reviews of these facts exist, including \cite{multibd}.

\subsection{The quotient MERA is the kinematic space of the two-sided black hole}
Let us compare the tensor network shown in Fig.~\ref{quotientTN} with the space of geodesics on a static slice of the two-sided BTZ black hole. 
Due to the discrete nature of the tensor network we may only quotient MERA by discrete scalings with $s = 2^k$ for $k \in \mathbb{Z}_+$. 

\begin{figure}
	\centering
	\includegraphics[height=0.35\textwidth]{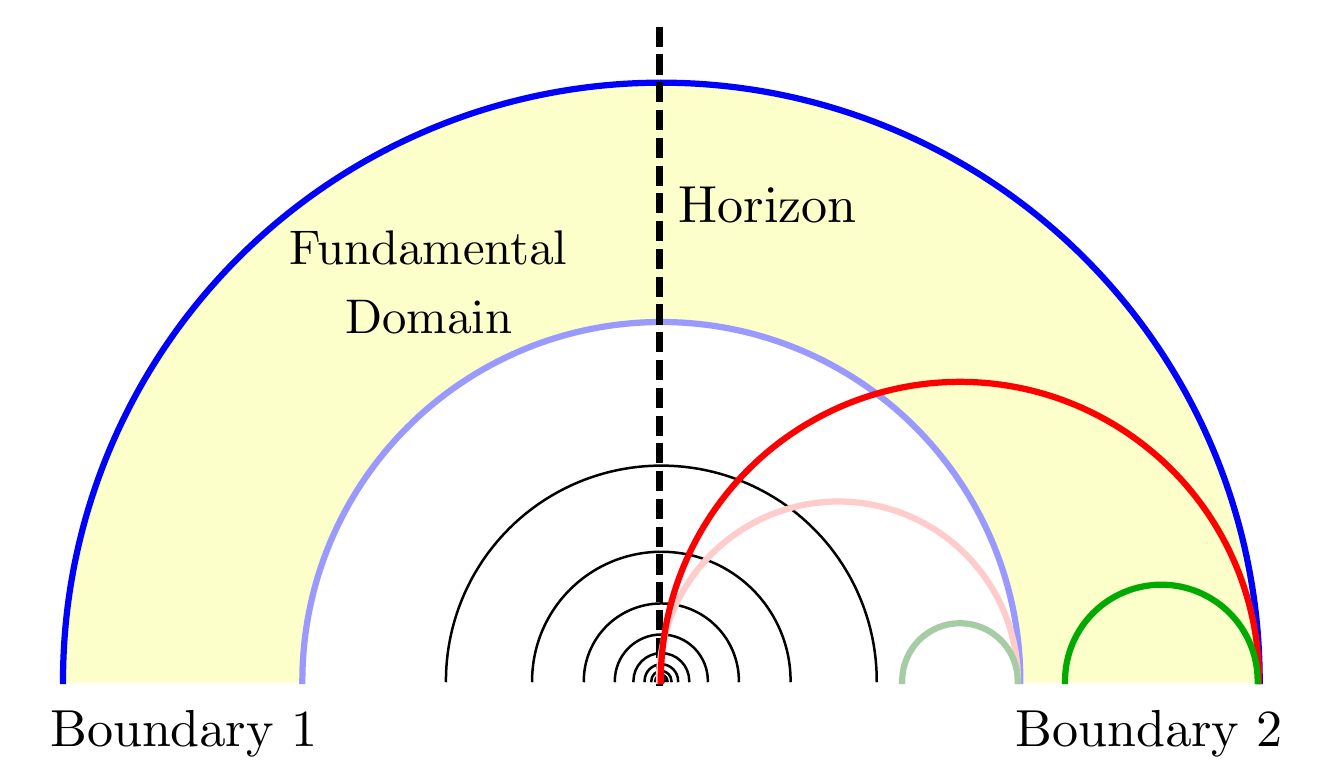}
	\hfill
	\includegraphics[height=0.35\textwidth]{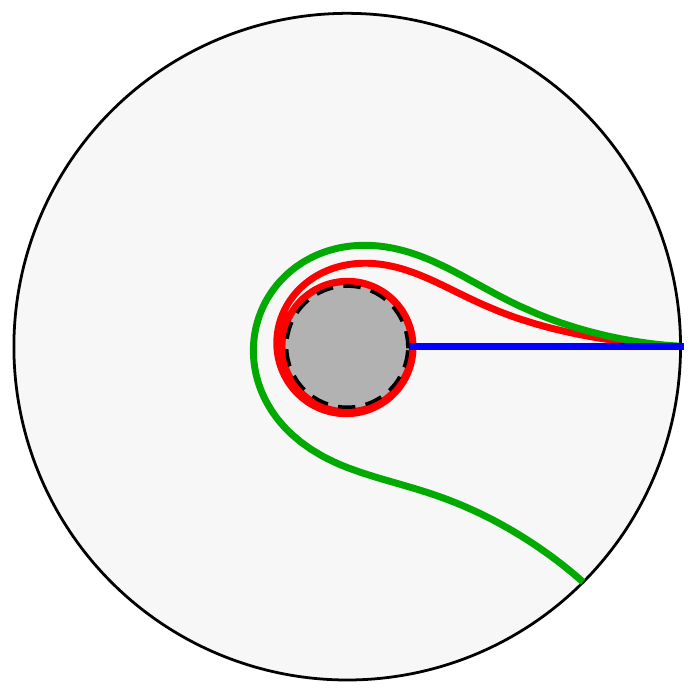}
	\caption{Identifying concentric geodesics on a static slice of AdS$_3$ produces a static slice of the two-sided BTZ black hole.  Three BTZ geodesics are shown, along with the AdS$_3$ geodesics from which they descend.  The green spacelike-identified geodesic is anchored on a single boundary.  The red lightlike-identified geodesic circles the horizon indefinitely. The blue timelike-identified geodesic connects the two sides.}
	\label{BTZidents}
	\end{figure}

\paragraph{Structure of the identifications} Observe that in the tensor network quotient in Fig.~\ref{quotientTN} not all identifications of indices are on the same footing. In the middle of the network we connect lines which are timelike-separated in the MERA sense. In the UV the identification joins indices that are spacelike-separated. The network contains two such regions, one on each side of the thermofield double. Separating the spacelike-identified regions from the timelike-identified one are single lines of tensors, which after the quotient form closed lightlike curves. These distinct components of the thermofield double MERA correspond to analogously distinguished classes of geodesics in the two-sided BTZ geometry.\footnote{They are also reminiscent of the `torus with whiskers' analyzed in \cite{funnytorus}.} We marked the three classes in Fig.~\ref{BTZidents}.

The first of these are the \emph{timelike-identified geodesics}. A canonical example of these are the geodesics in eq.~(\ref{identgeos}), which define the geometric quotient in Fig.~\ref{BTZidents}. More generally, a geodesic becomes identified with a timelike-separated image of itself if one of its endpoints is negative ($u < 0$) while the other one is positive ($v > 0$). This means that timelike-identified geodesics connect opposite sides of the two-sided black hole. They are horizon-crossing geodesics.

The \emph{spacelike-identified geodesics} remain on one side of the horizon. Their endpoints are either both positive ($0 < u < v$) or both negative ($u < v < 0$). In the bulk, such geodesics do not reach the horizon.

The marginal case separating the previous two are \emph{lightlike-identified geodesics}. Recall that $u$ and $v$, the left and right endpoint of a geodesic, are lightlike coordinates in kinematic space. Thus, the geodesic $(u,v)$ is lightlike-separated from its scaled image $(su, sv)$ if and only if $u = su = 0$ or $v = sv = 0$. This is consistent with the scope of the timelike-identified ($u < 0 < v$) and spacelike-identified regions ($u < v < 0$ and $0 < u < v$). 

The lightlike-identified geodesics are the borderline case, which separates horizon-crossing geodesics from those which remain a finite distance apart from the horizon. They are tangent to the horizon. A boundary-anchored geodesic can only become tangent to the horizon after spiraling around it infinitely many times. The infinite winding of the lightlike-identified geodesics can be seen in Fig.~\ref{BTZidents}. In the covering space, such geodesics cross infinitely many copies of one type of asymptotic boundary. For a more extensive discussion of infinitely winding geodesics in the BTZ geometry, consult \cite{lampros}.

\begin{figure}
	\centering
	\includegraphics[width=.7\textwidth]{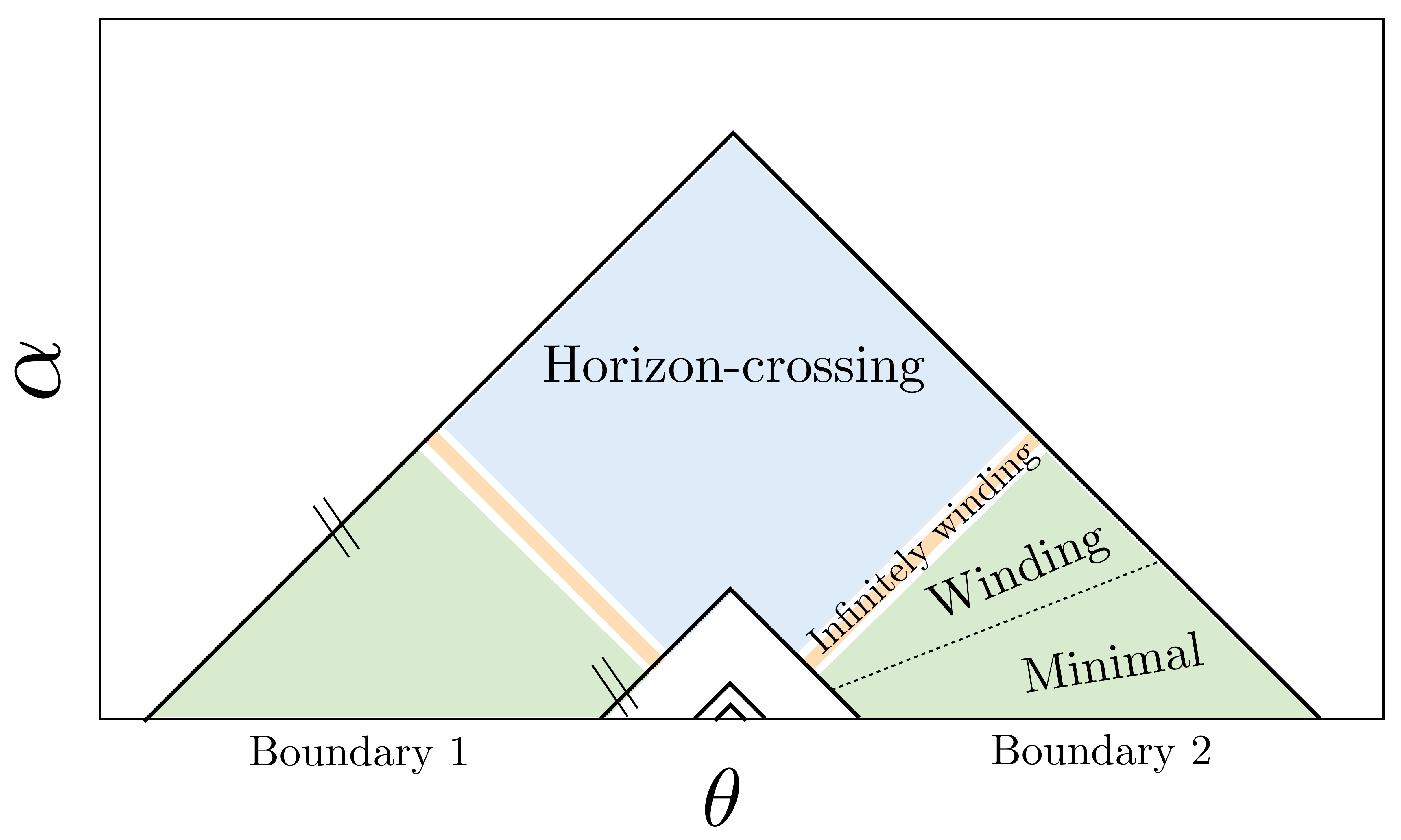}
	\caption{The kinematic space for the two-sided BTZ black hole, to be compared with Figs. \ref{quotientTN},\ref{BTZidents}.  The BTZ kinematic space is obtained as a quotient of the vacuum kinematic space, where two causal cuts are identified.  A fundamental domain is labeled above, which separates into horizon-crossing, winding (entwinement), and minimal (entropy) geodesics.  The points on the lightlike lines indicated correspond to infinitely winding geodesics.}
	\label{btzKS}
\end{figure}

\paragraph{Timelike-identified regions} In Ref.~\cite{quotientmera}, our co-authors and we explained that the timelike-identified region of the quotient network prepares the spectrum of the thermofield double state. In other words, this region is solely responsible for fixing the correlations between the two sides. This is exactly what we expect from the kinematic interpretation of MERA, which relates this region of the network to geodesics that cross the horizon and connect the two asymptotic boundaries. As an example, such geodesics were used to compute two-sided correlators in the thermofield double state in \cite{excursions}.

\paragraph{Black hole entropy} Following Fig.~\ref{fig:KSmutual}, the volume of the timelike-identified region in kinematic space computes the mutual information between the two sides:
\begin{equation}
I(L, R) = 2 S_{\rm BH}
\label{mithermal}
\end{equation}
Referring to the Crofton formula, this equation states that the area of the black hole horizon counts the geodesics that cross the horizon and connect the two sides of the wormhole geometry. 

It is instructive to recover this result explicitly in kinematic space by the use of the differential entropy formula. The latter asks for a complete set of geodesics tangent to the horizon, i.e. the lightlike-identified geodesics. Thus, the contour of integration is one full closed lightlike curve in kinematic space, for example $s u_0 < u \leq u_0$ and $v(u) = 0$. Substituting this into eq.~(\ref{sdiff}), we obtain:
\begin{equation}
S_{\rm BH} = 
- \int_{s\, u_0}^{u_0} du\,\,\frac{\partial S(u,0)}{\partial u}
= S(s\, u_0, 0) - S(u_0, 0) 
\label{sbhsdiff}
\end{equation}
We recognize this as the difference of the lengths of an $\infty$-wound geodesic and an `$(\infty-1)$-wound' geodesic. Indeed, taking the boundary-anchored endpoint of the geodesic from $u_0$ to $s u_0$ winds the already infinitely wound geodesic one additional time. The extra winding happens on the horizon of the black hole, which justifies eq.~(\ref{sbhsdiff}). Of course, the other closed lightlike-curve in the kinematic space of the BTZ black hole gives a similar result. There, we substitute geodesics $v_0 < v < s v_0$ and $u(v) = 0$ into the analogue of (\ref{sdiff}) appropriate for integrating over the $v$ coordinate:
\begin{equation}
S_{\rm diff} = 
\int_{v_0}^{s\, v_0} dv\,\,\frac{\partial S(u,v)}{\partial v} \Big|_{u = u(v)}
= S(0, s\, v_0) - S(0, v_0) = S_{\rm BH}
\label{sbhsdiff2}
\end{equation}

Interpreted in MERA, this computation recovers the minimal cut prescription in a novel setting: when the region whose entanglement we compute does not have endpoints. But as a bonus, we have explained why the thermal entropy can be read off from two different minimal cuts, on either side of the timelike-identified region. This is because we have two distinct families of lightlike-identified geodesics, which asymptote to the black hole horizon from either side of the wormhole. Their contributions add up to account for the factor of 2 in eq.~(\ref{mithermal}).

Furthermore, our network construction obtains the exact spectrum of the thermofield double state and not just the scaling of the entropy with the number of cuts. Indeed, in Ref.~\cite{quotientmera} our co-authors and we confirmed that the spectrum of the quotient network agrees quantitatively with the entanglement spectrum of the thermofield double state, including the numerical factors in eq.~(\ref{ltobeta}). This comparison was conducted in the critical Ising model, a decidedly non-holographic theory, in which case the kinematic space ought to be understood as the space of CFT intervals rather than the space of bulk geodesics.

\paragraph{Lightlike-identified geodesics and entwinement} Eqs.~(\ref{sbhsdiff}) and (\ref{sbhsdiff2}) involve geodesics, which wrap around the black hole horizon. These geodesics do not compute the entanglement entropy of any spatial interval in the thermofield double state. If we allow the use of their lengths in the differential entropy formula, however, we obtain correct geometric quantities, including some information-theoretically meaningful ones such as the entropy of the BTZ black hole \cite{lampros}. Emboldened by this, Ref.~\cite{entwinement} named a conjectured CFT avatar of the length of a non-minimal geodesic `entwinement.' Working in the conical defect geometry, the authors of \cite{entwinement} studied entwinement and concluded that it is related to entanglement among internal degrees of freedom. But an intrinsic definition of entwinement has remained an open question since then.

The quotient MERA manifests the relevance of entwinement in the CFT in the form of a lightlike-identified ray of tensors. Changing any one of these tensors affects the state on the entire CFT circle uniformly. Therefore, we may think of them as acting in the s-wave sector of the CFT, where no further spatial coarse-graining can be performed. Isometries in the lightlike-identified region separate the degrees of freedom which are internally entangled within the s-wave sector on one side from those which carry entanglement with the thermofield image. In the continuous geometry, entwinement is manifested by geodesics that wrap once or more around the black hole. Such `long geodesics' are sensitive to the internal organization of the CFT degrees of freedom, but also exhibit some degree of localization on the CFT circle. The discrete nature of MERA collapses the entire family of long geodesics into one line of tensors, which live on a lightlike-identified ray.

Entwinement is related to the structure of the CFT thermal state in the far infrared. At fixed temperature, it should therefore be more important for smaller circle sizes. By eq.~(\ref{ltobeta}) the effect of entwinement should wash out when $s \to \infty$ and gain in importance as $k = \log_2 s$ becomes of order 1. In Ref.~\cite{quotientmera} our co-authors and we confirmed these expectations. In particular, the lightlike-identified region in the quotient network is approximately isometric, with the approximation improving exponentially in $k$. For these reasons, we view the lightlike-identified regions in MERA as a tangible CFT realization of entwinement: because of their nearly isometric character and because their effect is completely delocalized in the CFT. 


\paragraph{Spacelike-identified regions} These prepare correlations between spatial regions on one side, just as they would in the vacuum. As unitary transformations between $\mathcal{H}_{\rm IR} \otimes \mathcal{H}_{\rm frozen}$ and $\mathcal{H}_{\rm UV}$, they select a local basis on each side in which the thermofield double state is presented. Different choices of local bases correspond to different conformal frames, distinguished by different UV cutoffs.

\section{Toward excited states: Geometry as compression}
\label{compressionsummary}
The intimate relation between integral geometry and information theory \cite{lastpaper} prompted us to look for an analogous structure on the CFT side of the holographic duality. We found it in the MERA tensor network. The two key properties, which MERA shares with kinematic space are the causal structure and the representation of entanglement entropy as a `flux' through a causal cut (counting lines in MERA and eq.~(\ref{eece}) in kinematic space). All our arguments originate from these two starting points.

We would like to extend our conclusions beyond the ground state MERA and its sub-networks---the cases discussed in Secs.~\ref{sec:therelation} and \ref{sec:BHmera}. We begin with the following observations:
\begin{enumerate}
\item In the ground state MERA, counting lines that cross a causal cut computes an entanglement entropy only in the optimized network. This feature is not a built-in property of the network; it is an emergent feature that arises after optimization and should not be expected to hold in excited states.  
\item The kinematic metric is not rigid. As seen in eq.~(\ref{kinmetric}), it depends on the state under consideration. When we consider excited states, the kinematic metric can only be relevant to the optimized network whose structure is adjusted for an efficient description of the state.
\end{enumerate}
In consequence, we must look for a flexible network that can incorporate the entanglement pattern in its structure. We propose a holographically motivated generalization of MERA which (a) shares the causal structure of MERA and kinematic space and (b) maintains the approximate relation between entanglement entropies and counting lines on causal cuts. Property (b) will require the state on a causal cut to be an approximate product state, a condition that is generally achievable only in holographic theories.

In information theoretic terms, such a tensor network is an \emph{iterative compression} algorithm: it maps the density matrix of every interval to a compressed state on its exclusive causal cone. Conditional mutual information---which for the ground state MERA was the number of isometries in a causal diamond---now provides a local \emph{density of compression}, namely the net reduction of the local Hilbert space dimension upon an isometric coarse-graining. Details of the construction, properties and limitations of the compression algorithm, as well as new insights about holographic geometries that follow from it, will be presented in an upcoming paper \cite{compression}.

\section{Summary}
\label{summary}

Holographic duality posits that gravity in an asymptotically anti-de Sitter space-time can be studied in a conformal field theory living on its asymptotic boundary. Of the many CFT quantities which encode the physics in AdS, entanglement entropies play a privileged role: they directly characterize the background space-time on top of which dynamics unfold \cite{rt1}. In the special case of the AdS$_3$/CFT$_2$ duality, where this characterization comes in the form of lengths of spacelike geodesics, it is particularly easy to convert it into a conventional picture of the bulk geometry in terms of points and distances. In \cite{lampros, lastpaper} we studied this conversion in detail and, in the process, discovered kinematic space.

From the CFT point of view, kinematic space is the geometry of intervals; in the bulk, it becomes the space of geodesics which is of interest in integral geometry. Kinematic space is a metric space of mixed signature: its causal structure reflects the containment relation among intervals while its volume form relates to variations of lengths of geodesics (eq.~\ref{kinvolume}). This last point is particularly significant in light of the Ryu-Takayanagi proposal: kinematic volumes compute conditional mutual informations of contiguous triples of intervals. The strikingly simple interpretation of kinematic space in information theory suggests that this concept will be useful for more general purposes. We will discuss one such use in an upcoming paper \cite{kinematicoperators}, which is concerned with defining a convenient `kinematic' basis of CFT operators that exploits the operator product expansion (OPE). 

\paragraph{Kinematic space and the MERA network} In the present paper, we report another application of kinematic space, which was independently discovered a decade ago in \cite{mera}. In order to write down the ground state wavefunction of a critical system in 1+1 dimensions, Vidal proposed MERA, a tensor network with a structure tailored to the scale dependence of entanglement entropies. The same dependence fixes the metric of kinematic space. This is more than a superficial similarity; we found that the properties of MERA are identical to the analogous properties of kinematic space, modulo the obvious limitations induced by discretization. The key common features of MERA and kinematic space are the following:

\begin{enumerate}
\item \emph{Causal propagation of information.} The locality and unitarity of the tensors comprising MERA implies that information propagates only within subregions dubbed 'causal cones.' This auxiliary notion of causality in MERA is identical to the causal structure of kinematic space, which encapsulates the containment relation among CFT intervals.


\item \emph{Localization of conditional mutual information.} In MERA, the conditional mutual information of three contiguous intervals localizes in a causal diamond. The same quantity defines the volume of a causal diamond in kinematic space. It is sensible to interpret conditional mutual information as a discrete volume form in MERA, because it simply counts isometries in a causal diamond. Using Stokes' theorem we can relate the number of isometries in a MERA region to the number of lines entering it. This is a MERA version of the differential entropy formula (eq.~\ref{sdiff}), a special case of which reproduces the commonly used cut counting prescription for estimating entanglement entropies.

\item \emph{Real space renormalization.} The MERA network performs RG transformations in real space and every spacelike cut defines a coarse-grained lattice. Kinematic space, understood as the space of CFT intervals, is a natural domain of real space cutoffs in the continuum. As in MERA, cutoff surfaces in kinematic space are not allowed to be timelike. An interesting type of cutoff is one that maximally coarse-grains two complementary intervals $A$ and $A^c$ (Sec. \ref{rgcausalcuts}). Both in MERA and in kinematic space such a cutoff retains a finite portion of the network (resp. space of geodesics), which prepares (resp. represents) the correlations between $A$ and $A^c$. This is in stark contrast to what a direct bulk interpretation of MERA would suggest.


\end{enumerate}

An important ingredient that underlies this detailed agreement between the kinematic geometry and MERA is the realization of entanglement entropies in the network as the size of causal cuts. This is a property that emerges in the optimized network for the vacuum but does not hold in general. Since it is a crucial prerequisite for the compatibility of the kinematic metric with the network, the simple MERA construction needs to be refined for general states. This can done by promoting features 1 and 2 above to principles for building the network. Adopting these principles leads to a representation of kinematic space as a compression algorithm, the details of which are reserved for a separate upcoming publication \cite{compression}.

\paragraph{Black Holes} Cutting the MERA network inhomogeneously effects local conformal transformations \cite{quotientmera}. In the kinematic proposal, this type of truncated MERA network discretizes the space of geodesics of a locally AdS$_3$ geometry. We discussed these kinematic spaces in Sec.~\ref{sec:BHmera}, with a particular emphasis on the BTZ black hole which involves an additional, non-trivial quotient. Both the thermofield double MERA and the BTZ kinematic space divide into three regions, distinguished by the timelike versus spacelike action of the quotient.

The agreement of MERA and kinematic space goes beyond this structural similarity. The timelike-identified region of MERA, which relates to geodesics that connect the two asymptotic boundaries, is responsible for preparing the correlations (entanglement spectrum) between the two sides. The thermal entropy of each asymptotic boundary equals half the volume of this timelike-identified region, which equates the black hole entropy with the total number of geodesics connecting the two sides. On the other hand, the spacelike-identified sectors of the network, which correspond to geodesics that stay on one side of the black hole, prepare spatial correlations within each asymptotic region.

An intriguing part of the thermofield double MERA are two single lines of tensors on which the quotient acts in a lightlike manner. Geometrically, they relate to geodesics which asymptote to the black hole horizon after winding around it infinitely many times. These tensors do not perform any type of coarse-graining of local degrees of freedom, just as winding geodesics do not compute entanglement entropies of localized intervals. We view these tensors as a concrete realization of the concept of entwinement---entanglement of internal or gauged degrees of freedom, which is conjecturally related to the length of non-minimal geodesics \cite{entwinement}.

\paragraph{Comparison to other tensor network proposals} 
Recent work on applications of tensor networks to holography includes novel networks distinct from MERA, which discretize the spatial geometry of anti-de Sitter space directly \cite{errorcorrectingnetwork, patrickmichael}. At present, it is unclear whether these networks can prepare the wavefunction of the CFT ground state. Likewise, the question of how such networks can be extended to encode wavefunctions of excited states and their dual geometries remains open. On the other hand, the AdS/CFT correspondence essentially guarantees that a network of this type---one whose structure mimics the fabric of the bulk geometry---must exist. After all, the change of basis that takes CFT degrees of freedom into low energy effective fields in the bulk can always be presented in the form of a tensor network.

While the quest for such a network continues, we believe that it is also important to work with those tensor networks, which are known to prepare the correct CFT states. No ansatz is successful by accident; if MERA prepares the CFT ground state, it does so because its design was guided by the correct set of principles. As we explained above, the principles underlying MERA are also the defining features of kinematic space, which in holographic theories becomes the space of bulk geodesics. In this way, MERA makes contact with geometric concepts without any input from the bulk side; it is a showcase example of an \emph{emergent} geometry. At the same time, because MERA requires no input from the bulk and because kinematic space can be merely a space of CFT intervals (rather than a space of geodesics), our analysis applies to non-holographic theories. Ref.~\cite{quotientmera} illustrates that this approach can benefit the more traditional, CFT-centered applications of tensor networks.

\section*{Acknowledgments}
We thank Glen Evenbly, Patrick Hayden, Esperanza Lopez, Don Marolf, Rob Myers, John Preskill, Xiao-Liang Qi, Joan Sim{\'o}n, Leonard Susskind, Brian Swingle and Guifr{\'e} Vidal for useful discussions. BC, SM and JS thank Caltech, BC thanks the University of Amsterdam and the University of Edinburgh, and JS thanks Princeton University for hospitality. BC thanks the organizers of ``Holographic duality for condensed matter physics'' and KITPC-CAS in Beijing. We all thank the organizers of ``Quantum Gravity Foundations: UV to IR,'' ``Closing the Entanglement Gap: Quantum Information, Quantum Matter and Quantum Fields,'' and the Follow-On Program held at KITP (supported in part by the National Science Foundation under Grant No. NSF PHY11-25915), and of ``Quantum Information Theory in Quantum Gravity II'' meeting held at the Perimeter Institute for Theoretical Physics (supported by the Government of Canada through Industry Canada and by the Province of Ontario through the Ministry of Research and Innovation). BC and JS thank the organizers of the workshop ``AdS/CFT and Quantum Gravity'' at Centre de Recherches Math{\'e}matiques at the University of Montreal. SM was supported in part by an award from the Department of Energy (DOE) Office of Science Graduate Fellowship Program.

%

\end{document}